\documentclass[nofootinbib,reprint,amsmath,amssymb,showkeys,superscriptaddress,showpacs,aps]{revtex4-1}

\usepackage[normalem]{ulem}
\usepackage{graphicx}
\usepackage{dcolumn}
\usepackage{xcolor,colortbl}
\usepackage{multirow}
\usepackage{bm}

\usepackage{hyperref}
\hypersetup{colorlinks, linkcolor={red},citecolor={blue},urlcolor={blue}}

\usepackage{accents}
\usepackage{scalerel}
\usepackage{tabularx,siunitx}
\usepackage{wasysym}
\usepackage{mathrsfs}
\usepackage{comment}
\usepackage{amsfonts}
\usepackage{enumitem}
\usepackage{amsmath,amssymb}

\begin{document}

\title[Special issue EPJ plus]{An empirical investigation into cosmological tensions.}

\author{{Ramon}~{de Sá}}
\email{desaramon4@gmail.com}
\affiliation{Instituto de Fisica, Universidade Federal Fluminense, 24210-346 Niteroi, RJ, Brazil}

\author{{Micol}~{Benetti}}
\email{micol.benetti@unina.it}
\affiliation{Scuola Superiore Meridionale,  Largo San Marcellino 10, 80138 Napoli, Italy}
\affiliation{Istituto Nazionale di Fisica Nucleare, Sezione di Napoli, Via Cintia 80126 Napoli, Italy}

\author{{Leila}~{Graef}}
\email{leilagraef@id.uff.br}
\affiliation{Instituto de Fisica, Universidade Federal Fluminense, 24210-346 Niteroi, RJ, Brazil}

\date{\today}

\begin{abstract}
The possibility that the $H_0$ tension is a sign of a physics beyond the $\Lambda$CDM model is one of the most exciting possibilities in modern cosmology.
The challenge of solving this problem is complicated by several factors, including the worsening of the tension on $\sigma_8$ parameter when that on $H_0$ is raised. 
Furthermore, the perspective from which the problem is viewed should not be underestimated, since the tension on $H_0$ can also be interpreted as a tension on the size of the acoustic horizon, $r_s$, which deserves proper discussion. 
The common approach in the literature consists in proposing a new model that can resolve the tension and treat the new parameters of the theory as free in the analysis. However, allowing additional parameters to vary  often results in larger uncertainties on the inferred cosmological parameters, causing an apparent relaxing in the tension due to the broaden in the posterior, instead of a genuine shift in the central value of $H_{0}$. To avoid this, we consider here an empirical approach that assumes specific non-standard values of the $\Lambda$CDM extensions and we analyze how the important parameters in the context of the tension vary accordingly. 
For our purposes, we study simple extensions of the standard cosmological model, such as a phantom DE component (with Equation of State $w < -1$) and extra relativistic species in the early universe (so that $N_{eff} > 3.046$). 
We obtain relations between variation in the value of $w$ and $N_{eff}$ and changes in $H_0$,  $r_s$ and $\sigma_8$. 
In this way an empirical relation between $H_0$ and $\sigma_8$ is provided, that is a first step in understanding  which classes of theoretical models, and with which characteristics, could be able to break the correlation between the two tensions. 
\end{abstract}


\maketitle
\section{Introduction}
\label{sec:Introduction}
In the cosmological standard approach, the Cosmic Microwave Background
(CMB) provides the initial condition for the further evolution of structures in
the universe. The standard  model of cosmology, the $\Lambda$CDM ($\Lambda$
Cold Dark Matter), is then used to get predictions of large-scale structure 
 of the late-time universe, whose observation  can provide an end-to-end test of the theory. However, we have signs that this approach can be failing. One  evidence is associated to the so called
 Hubble  tension,  the discrepancy in the Hubble constant value, $H_0$, when measured by direct observations or inferred by the theory. Currently the tension between the early (model dependent) and late time constraints on $H_{0}$ is more than 4$\sigma$ (and less than 6$\sigma$) \cite{Planck:2018vyg, Riess:2021jrx}. Several possibilities have been investigated, like varying the number of neutrinos species \cite{Planck:2018vyg,Benetti:2017gvm,Benetti:2017juy},  considering both a closed and an open universe \cite{Vagnozzi:2020rcz, DiValentino:2020hov, Bose:2020cjb,Gonzalez:2021ojp,Efstathiou:2020wem,Vagnozzi:2020dfn} or also reformulating
the $H_0$ tension as a tension in the CMB monopole temperature $T_{0}$ \cite{Ivanov:2020mfr}. 
More in general, a plethora of proposals to alleviate the tension have been studied and here we can roughly classify them into few categories \cite{DiValentino:2021izs} as the
early dark
energy  \cite{Herold:2021ksg, Reeves:2022aoi, Karwal:2016vyq,Vagnozzi:2021gjh} and late dark energy \cite{Zhao:2017cud} models, interacting dark energy mechanisms \cite{Bolotin:2013jpa,Wang:2016lxa,vonMarttens:2020apn,vonMarttens:2022xyr,Benetti:2019lxu,Salzano:2021zxk,Benetti:2021div,DiValentino:2019ffd}, models with extra relativistic degrees of freedom \cite{Kuroyanagi:2020sfw, Domenech:2020kqm, Vagnozzi:2020gtf, Sugiyama:2020roc, Namba:2020kij, Graef:2018fzu, Kuroyanagi:2014nba}, models with extra interactions \cite{Aloni:2021eaq, Joseph:2022jsf,Schoneberg:2022grr}, unified cosmologies \cite{Hova:2010na}, modified gravity \cite{Braglia:2020iik,Raveri:2019mxg,Horndeski:1974wa,deBrito:2020xhy,Frusciante:2020gkx,Benetti:2020hxp,Benetti:2018zhv}, non-standard inflationary models \cite{Benetti:2017juy, Rodrigues:2021txa},
modified recombination history \cite{Chiang:2018xpn}, physics of the critical phenomena \cite{Banihashemi:2018oxo}, and alternative proposals \cite{DiValentino:2021izs,Spallicci:2021kye,Capozziello:2020nyq, Benetti:2019gmo}. 
{
Furthermore, it should not be underestimated that the tensions that emerged are between $H_0$ estimates obtained in very different ways. While on the one hand there is a direct measure of $H_0$ from the observed recessional velocity of galaxies, i.e. a \textit{kinetic measure} of Hubble constant  \cite{Freedman:2020dne}, on the other hand there is an inference of $H_0$ from a considered model that assumes that the universe is composed of particular densities of matter and energy. That is, a \textit{dynamic measure} of Hubble constant. Finally, it should not be forgotten that the estimate of $H_0$ from kinematic measurements is not calculated exactly at $z=0$ but is inferred from observations at slightly higher reshifts, i.e $z > 0.02$, with the assumption that there is no particular local physics that would change such an estimate (for a proposal of late dark energy transitions, see Ref. \cite{Benevento:2020fev}). In any case, assuming that the tension is real, it must be emphasised that usually} the proposed models are studied allowing additional parameters of the standard theory to vary. Such procedure often  results in larger uncertainties
on the inferred cosmological parameters, including $H_0$.
This happens since marginalizing over additional parameters implies in broadening the posterior of all cosmological parameters, especially the ones strongly correlated  with the new parameters. Consequently,  the Hubble tension can be relaxed simply because of an increase in the uncertainty in the $H_0$ value inferred from CMB data, and not due to a true shift in its central value, as discussed in \cite{DiValentino:2021izs}.  However, this is not the case when a certain model predicts (or fix) a non-standard value for the new parameter of the theory. In this case, the degrees of freedom of
the model are the same of the standard one, helping to understand whether the model really relieve the $H_0$ tension problem. 

Among the several possibilities, in this work we focus in two of the simplest extensions of the standard model. The first one consists in  considering a  time-varying dark energy. As its well known, in a $\Lambda$CDM model dark energy is described by a cosmological constant. Nevertheless  there are several well motivated  reasons to consider  that dark energy may evolve in time. A time-dependent quintessence-like dark energy may affect the status of the tensions between local and indirect  measurements of cosmological parameters \footnote{In the case of quintessence, \cite{Banerjee:2020xcn}, or any late-time dark energy model with an equation of state $w > -1$  \cite{Lee:2022cyh},  one can prove that $H_0$ is always sent to lower values, implying in  severe constraints to such models.}. A second  simple possibility that we are going to consider here involves the presence of extra relativistic species in the early universe, so that $N_{eff} > 3.046$, where $N_{eff}$ is the effective number of relativistic species. Indeed, several particle physics theories can justify an extra contribution, and cosmological data do not exclude a small deviation from the value of three neutrino families \cite{Planck:2018vyg}. 
Recently, for these models the “sweet spot” between decrease in Bayesian evidence and reduction in the $H_0$ tension was shown, i.e. values of $w$ and $N_{eff}$ which, if predicted by a model, would lead to an alleviation in the $H_0$ tension while  not being strongly disfavoured with respect to $\Lambda$CDM \cite{Vagnozzi:2019ezj}. It was found that models with $N_{eff} = 3.95$ or $w = - 1.3$ solve the $H_0$ tension but are strongly disfavored by the data. On the other hand, models with $N_{eff} = 3.45$ or $w = - 1.1$ brings the tension down to 1.5$\sigma$ and 2.0$\sigma$ respectively, but are weakly and strongly disfavored with respect to $\Lambda$CDM. 
In our work, we go beyond this study addressing not only the $H_0$ tension but also the one in the value of the clustering parameter, $\sigma_{8}$. Indeed, it takes different values depending on the data considered, and this tension can be related to the $H_{0}$ problem \footnote{Note that $H_0$ is a fitting parameter in the flat $\Lambda$CDM model and one can  assume it may be not a constant (see, for example,  Refs. \cite{Krishnan:2020vaf, Colgain:2022rxy, Colgain:2022nlb} for the  possibility of evolution of $H_{0}$ in the late universe). If $H_0$ evolves in the late universe, so too must $\Omega_{m0}$, in which case one expects $S_{8}$ $\propto \sqrt{\Omega_{m0}}$ to also be impacted.}.  It has been shown in the literature that many of the attempts to alleviate the $H_0$ problem with SH$0$ES data \cite{Riess:2021jrx} tend to worsen the $\sigma_{8}$ one with lensing observations \cite{Jedamzik:2020zmd}, and the opposite is also true \footnote{Possible exceptions can be found, for example, in the context of interacting dark matter/dark energy models and models of dark matter with evolving, and negative, equation-of-state - see for example Ref. \cite{Naidoo:2022rda}.}. Since the new physics that increases the current expansion rate usually suppresses the structure formation, this effect tend to be compensated by an increasing in the density of the cold dark matter, which leads to an increase in the $\sigma_{8}$ tension. This one has become statistically significant and it is now at the level of 3.1s with KiDS-100 \cite{Heymans:2020gsg} and 2.5s with DES-Y3 \cite{DES:2021bvc}, while Planck is preferring a higher value \cite{Planck:2018vyg} \footnote{See Ref. \cite{Nunes:2021ipq} for a different perspective on the statistical significance of the  $\sigma_{8}$ tension}. 
So its crucial to understand possible ways to break the correlation between these effects.

In this work we focus specifically in obtaining empirical relation between these parameters. By comparing these relations we can better understand how their effects interplay driving  a correlation between the tensions in $H_0$ and $\sigma_8$. Also, another important issue we consider here is that the Hubble constant and the sound horizon, $r_s$,  are closely related \cite{Zhang:2020uan,Bernal:2016gxb,Arendse:2019hev} and link the late-time and early time cosmology. 
 While reducing the sound horizon  can reduce the Hubble tension from $\sim 4\sigma$  to $\sim 2\sigma$, one cannot bring CMB into a full agreement with SH$0$ES by a reduction of $r_s$ alone without running into trouble with either the Baryonic Acoustic Oscillation (BAO) or the galaxy weak lensing data. 
 So, we extend the previous analysis of Ref. \cite{Vagnozzi:2019ezj} to include also an $r_{s}$ analysis.   
 
This work intends to be a guide for theoretical model buildings by providing  model independent relations between the parameters that plays important roles in the tension problem. While no	simple 	extension of $\Lambda$CDM	can	solve	all	tensions	and	accommodate	all	data, it is  important to  understand  how $H_0$ and $\sigma_8$ should scale with the  extra parameters in order to break their correlated  behaviour  in any  desired model. Providing  a better understanding  on the main relations that  a desired  model   should satisfy can be very useful for theoreticals to construct more fundamental  models that could show a different behaviour with regard to those  tensions.

\section{$\Lambda$CDM extensions}
\label{sec:Models}

We choose to apply our strategy to two extensions of the $\Lambda$CDM model well studied in the literature, nominally producing changes in the physics of the universe both at large and small scales. 
We shall therefore consider a model in which the dark energy equation of state is always constant with redshift, but takes on values other than the canonical $w=-1$. This is justified by several observations using both primordial and recent universe data, in which a free $w$ parameter generally prefers larger negative values \cite{Planck:2018vyg, Bargiacchi:2021hdp,Joudaki:2016kym,Zhang:2018air,Visinelli:2019qqu,DiValentino:2019jae, Keeley:2019esp}. 
At the same time,
there are several mechanisms which can motivate higher values for $N_{eff}$ (for a complete discussion, see the review \cite{Abdalla:2022yfr} and therein references). In addition to the possibility of existence of extra relativistic neutrinos from particle physics, other examples are the early universe phenomena which predicts a significant stochastic background of gravitational waves. Indeed, the early universe models that predicts a blue tilt in the primordial tensor spectrum, among several other possibilities, can increase the value of $N_{eff}$ of approximately $\sim 0.10$ \cite{Kuroyanagi:2020sfw,Domenech:2020kqm,Vagnozzi:2020gtf,Sugiyama:2020roc,Namba:2020kij,Graef:2018fzu,Kuroyanagi:2014nba}\footnote{Note that models with  high positive values of the tensor tilt ($n_{T} \sim 1$) must include  some mechanism to change the power-law form of the spectrum from blue to red at some frequency, in order to ensure consistency with BBN and LIGO upper limits on the  Stochastic Background of Gravitational Waves \cite{Benetti:2021uea}.}. 

As mentioned before, although no simple mechanism can alleviate both  
tensions, its important to better understand how each (late and early time) mechanisms are related and how they impact the parameters $H_{0}$, $r_s$ and $\sigma_{8}$. 
Assuming a flat universe geometry, the Hubble parameter $H(z)$ can be related to the effective number of relativistic species $N_{eff}$ through the expression below (see Ref. \cite{Maggiore:1999vm} for more details),

\begin{equation}\label{eq:early}
    \frac{H(z)}{H_{0}}= \sqrt{\Omega_{m}(1+z)^{3}+ \Omega_{\gamma}(1+0.2271 \; N_{eff})(1+z)^{4} + \Omega_{\Lambda}},
\end{equation}
where $H_0$ is the current value of the Hubble parameter, the total matter density, $\Omega_{m}$, is the sum of $\Omega_{cdm}$ and $\Omega_{b}$, the cold dark matter and baryons density, respectively; $\Omega_{\gamma}$ stands for the photons density (essentially fixed by the temperature of the CMB) while $\Omega_{\Lambda}$ indicates the cosmological constant.

Concerning the late time universe, one can easily see the connection between the Hubble parameter and the equation of state of the current dominant component of the universe by writing the background evolution as follows,

\begin{equation}\label{eq:late}
       \frac{H(z)}{H_{0}}= \sqrt{\Omega_{m}(1+z)^{3}+ \Omega_{r}(1+z)^{4}+ \Omega_{\Lambda}(1+z)^{3(1+w)}},
\end{equation}
where $w$ is the equation of state of the dominant component of the late-time universe, which is considered to be the dark energy. 
Note that we are assuming the flatness of the universe, so that $\Omega_{\Lambda}= 1 - \Omega_{m}-\Omega_{r}$. Also, an EoS of $w=-1$ recover a cosmological constant, i.e independent with redshift.

The $H_{0}$ tension can also be thought of from the perspective of a mismatch in the sound horizon at the last scattering, $r_{s}$. 
Indeed, the SH$0$ES estimates combined with BAO+Pantheon data need $r_s \sim 137$ Mpc, while Planck data assuming $\Lambda$CDM model indicates $r_s \sim 147$ Mpc. This means that a $10$ Mpc difference ($\sim 5\%$) in the estimate of the sound horizon at the last scattering can resolve the tension in the expansion rate of the universe today \cite{Bernal:2016gxb}.

The sound horizon at the last scattering (with redshift $z_{*}$) is given by the following expression,

\begin{equation}
    r_{s}= \int^{\infty}_{z_{*}} dz \; \frac{c_{s}(z)}{H(z)}
\end{equation}

where we use $r_s (z_{*}) = r_s$  for the sake of simplicity.

While the constrained angular size of the sound horizon, $\theta_{s}$ is given by, 
\begin{equation}
    \theta_{s} = \frac{r_s}{D_{A}(z_{*})},
\end{equation}

where the angular diameter distance, $D_{A}$,  is a model dependent quantity, as 

\begin{equation}
 D_{A}(z_{*})= \frac{1}{1+z_{*}}\int^{z_{*}}_{0} dz \; \frac{1}{H(z)}.   
\end{equation}

The smaller angular size of the sound horizon is, the more the CMB peaks are shifted to the smaller scales (larger multipoles). This can be brought in accordance with the data
either by increasing the early-time expansion rate, i.e via an $N_{eff}$ extra contribution
\cite{Arendse:2019hev}, or modifying the current expansion rate \cite{Knox:2019rjx}. 
In the first case, the increase in the number of relativistic species at the recombination time is able to both increasing the value of $H_0$ and decreasing the value of $r_s$, thus moving the two quantities in the right direction to relax the tension (but not enough to solve it) \cite{Zhang:2020uan,Bernal:2016gxb,Arendse:2019hev}.
In the second case, instead,  $D_{A}(z_{*})$ value decreases recovering an higher $\theta_{s}$ consistently with that predicted by Planck, while $r_s$ is not significantly affected.

Furthermore, we should note that letting  both $N_{eff}$ and $w$ values free to vary with respect to the standard $\Lambda$CDM values leaves the constraints of these parameters always compatible with those of the standard model, and the alleviation of the tension is mainly due to a volume effect due to the increase of the parameter error bars \cite{Arendse:2019hev}. Hereafter, we overcome this effect by fixing the parameters at certain values rather than leaving them free to vary. This will allow us to investigate the relationships that link these parameters to the tensions resolution without necessarily expanding the volume of the parametric space as increasing errors. We explain in the following the empirical analysis we perform to further investigate these relations.

\section{Methodology}\label{sec:Method}

In our analysis we consider two model, namely the $\overline{w}CDM$ and $\overline{N}CDM$, where the over bar stands for the specific extension of the standard cosmological model we are considering. Indeed, in the first case we assume a background evolution as Eq.(\ref{eq:late}), with $w$ the equation of state of dark energy fixed at several values, i.e $-1.3$, $-1.2$, $-1.1$ and $-1$. In the $\overline{N}CDM$ model, the extension refers to the $N_{eff}$ parameter, that is analysed using the values of $3.046$, $3.15$, $3.55$ and $3.95$ with a background evolution as Eq.(\ref{eq:early}). The values assumed are compatible (exceeding only by one extreme case for $N_{eff}=3.95$) with observational data (using Planck TT+lowE \cite{Planck:2018vyg}), since Planck constrains in this data analysis $w=-1.57^{+0.60}_{-0.48}$ and $N_{eff}=3.00^{+0.57}_{-0.53}$ at $95\%$.

So, we build the two $\Lambda$CDM extensions varying the usual cosmological parameters, namely, the physical baryon density, $\Omega_b h^2$, the physical
cold dark matter density, $\Omega_{cdm} h^2$, the ratio between the sound horizon and the angular diameter distance at decoupling,
$\theta_s$, the optical depth $\tau$, the primordial amplitude, $A_s$ and the spectral index $n_s$. But unlike the standard model, we set the value of $w$ and $N_{eff}$ at  values different from $-1$ and $3.046$. We consider purely adiabatic initial conditions; the sum of neutrino masses is fixed to 0.06 eV, and we limit the analysis to scalar perturbations with $k^*=0.05$ Mpc. For cosmological parameters, we assume the usual broad and linear prior as in the standard cosmological model analysis.

We use the a Monte Carlo Markov chain exploration of the parameters space using the available package CosmoMC \cite{Lewis:2002ah}. We then constraint the free parameters of the theory and obtain the derived parameters of the model, such as the Hubble constant today, $H_0$, the amplitude of the matter power spectrum at $8$ Mpc, $\sigma_8$, and the sound horizon at last scattering, $r_s$.

We consider an extended dataset comprising Cosmic Microwave Background measurements, through the Planck (2018) likelihoods \cite{Planck:2019nip}  \footnote{ We use Planck likelihood  ``TT,TE,EE+lowE" by combination of temperature TT, polarization EE and their cross-correlation TE power spectra  over the range $\ell \in [30, 2508]$, the low-$\ell$ temperature Commander likelihood, and the low-$\ell$ SimAll EE likelihood.}, the CMB lensing reconstruction power spectrum~\cite{Planck:2019nip,Planck:2018lbu}, the Baryon Acoustic Oscillation measurements from 6dFGS~\cite{Beutler:2011hx}, SDSS-MGS~\cite{Ross:2014qpa}, and BOSS DR12~\cite{BOSS:2016wmc} surveys, the type Ia SN Pantheon compilation~\cite{Scolnic:2017caz}.

In the next section, we show the explored planes $w-H_0$, $N_{eff}-H_0$, $w-\sigma_8$, $N_{eff}-\sigma_8$, $w-r_s$, $N_{eff}-r_s$, interpolating the points obtained varying the values of $w$ and $N_eff$ and then  finding the best fit in the linear correlation. 
In the end we also analyse the inferred $H_0$ in the model $\overline{w N}$CDM, where both $w$ and $N_{eff}$ assume values different from the $\Lambda$CDM model.

\section{Results}\label{sec:Results}

In this section we present and discuss the results of our analysis. The constrained values of the parameters of interest,  $H_{0}$, $\sigma_{8}$ and $r_{s}$ are listed in Tab.\ref{tab1}, where also the reference $\Lambda$CDM model, with fixed $w=-1$ and $N_{eff}=3.046$, is quoted.

\begin{table}[h]
\begin{center}
\caption{Parameter constraints in the $\overline{w}CDM$ and $\overline{N}CDM$ models. The first column shows the values of $w$ and $N_{eff}$ set for the two extended models. All the value are in $68\%$ C.L. Below $H_{0}$ is shown in units  km/s/Mpc, $r_s$ is in Mpc and $\sigma_8$ is the adimensional amplitude of mass fluctuations on scales of $8~h^{-1}$ Mpc.}
\label{tab1}
\begin{tabular}{@{}llll@{}}
\toprule
$\Lambda$CDM & $H_0$  & $\sigma_8$  & $rs$\\
& $67.72 \pm 0.41$   & $0.8099 \pm 0.0059$   & $144.59  \pm 0.21$  \\
\toprule
$\overline{w}$CDM & $H_0$  & $\sigma_8$  & $rs$\\
$w=-1.1$    &  $69.93 \pm  0.45$   & $0.8391 \pm 0.0060$  & $144.26 \pm  0.21$   \\
$w=-1.2$    &  $72.00 \pm  0.50$   &  $0.8663 \pm 0.0063$   & $143.94 \pm 0.20$  \\
$w=-1.3$    & $73.92 \pm  0.56$   & $0.8914 \pm 0.0066$   & $143.61 \pm 0.21$  \\
\toprule
$\overline{N}$CDM & $H_0$  & $\sigma_8$  & $rs$\\
$N_{eff}=3.15$    & $68.36 \pm 0.42$   &  $0.8149 \pm 0.0059$   & $143.59 \pm 0.22$   \\
$N_{eff}=3.55$    & $70.76 \pm 0.42$   & $0.8330 \pm 0.0061$  & $139.95 \pm  0.21$  \\
$N_{eff}=3.95$    & $73.11 \pm 0.46$   & $0.8499 \pm 0.0067$  & $136.54 \pm  0.20$  \\
\botrule
\end{tabular}
\end{center}
\end{table}

In Figs.\ref{fig:grafw} - \ref{fig:rsneff} we show the normalized probability distribution function for the parameter $H_{0}$, $\sigma$ and $r_{s}$ in each model considered. Using their best fit values, we can relate $\Delta w$ and $\Delta N_{eff}$ with the inferred values for $\Delta H_{0}$, $\Delta \sigma_{8}$ and $\Delta r_{s}$, where the $\Delta$ indicates the difference with respect to the standard model values. 
This analysis allows us to construct an empirical relationship between $\Delta H_{0}$ and $\Delta \sigma_{8}$, allowing us to discuss the correlation between the two tensions. 

Below we start by showing the results for the $H_{0}$ parameter in each of the extended models considered. 

\subsection{Results on $\Delta H_{0}$}\label{subsec1}

Here we show the relations obtained between the  inferred values of $H_0$ and the variation of $w$ and $N_{eff}$ in the extended models considered. We begin by considering the results for  $\overline{w}$CDM models, 
 and then for  $\overline{N}$CDM models.
  The results are compared with the local measurements by Cepheids, highlighted in the figures as a green column, which corresponds to the value $H_0 = 73.24 \pm 1.74 $ km/s/Mpc \cite{Riess:2018uxu}. 

\begin{figure}[t]
    \centering
    \includegraphics[scale=0.6]{ 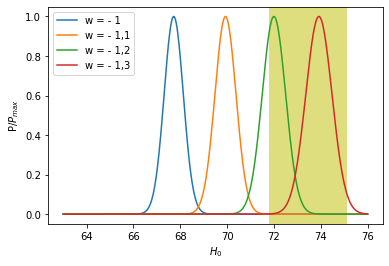}
    \caption{Normalized posterior distributions of $H_0$ for different fixed values of $w$. The green shaded region is the 1$\sigma$ region for $H_0$ determined by the Hubble Space Telescope and the SHOES Team, yielding $H_0 = 73.24 \pm 1.74$ km/s/Mpc \cite{Riess:2018uxu}.}
    
    \label{fig:grafw}
\end{figure}

In Figure \ref{fig:grafw}, we show the normalized probability distribution function for the selected models with different values for $w$. We show the result for the standard model $w =-1$ (blue line), and for the extended modes with $w = -1.1$ (orange line), $w = -1.2$  (green line) and $w = -1.3$ (red line). 
We can note that for $w=-1.3$, the current Hubble constant is $H_0 = 73.92 \pm 0.56$  km/s/Mpc, close to the value inferred by local measurements, but also the prediction for the lower value of $w=-1.2$ is in agreement with the Cepheids estimation.

Considering the best fit values obtained for these parameters in each model, we can relate $\Delta w$  with the inferred values for $\Delta H_{0}$ by adjusting the best curve to fit the data. This provides an approximated linear relation,

\begin{equation}
    \Delta H_0 \equiv H_0 - H_0^{\Lambda CDM} \approx -20.6(1 + w) \equiv -20.6 \Delta w.
    \label{deltaw}
\end{equation}

\begin{figure}[t]
    \centering
    \includegraphics[scale=0.6]{ 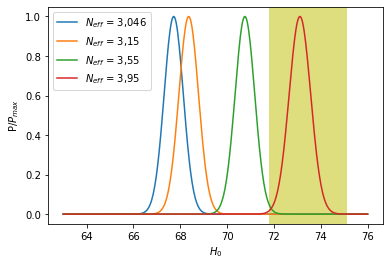}
    \caption{Normalized posterior distributions of $H_0$ for different choices of $N_{eff}$. The green shaded region is the 1$\sigma$ region for $H_0$ determined by the Hubble Space Telescope and the SHOES  Team, yielding $H_0 = 73.24 \pm 1.74$ km/s/Mpc \cite{Riess:2018uxu}.}
    
    \label{fig:grafneff}
\end{figure}

In the same way, we show the normalized probability distribution function for the selected models with different values for $N_{eff}$ in Figure \ref{fig:grafneff}. 
We achieve the concordance with SH$0$ES values by assuming an $N_{eff} = 3.95$, that allows an $H_0 = 73.11 \pm 0.46$ km/s/Mpc. %
This $N_{eff}$ value is outside the constraints allowed by cosmological data and has been taken here only for illustrative purposes. The most reasonable value on $N_{eff}=3.55$ show a 3$\sigma$ agreement with the predictions of Riess \textit{et al.}.

In this case, relating the $\Delta N_{eff}$ with the inferred values for $\Delta H_{0}$ using the best fit of our analysis we found the linear expression,

\begin{equation}
    \Delta H_0 \equiv H_0 - H_0^{\Lambda CDM} \approx 5.9 (N_{eff} - 3.046) \equiv 5.9 \Delta N_{eff}.
    \label{novonef}
\end{equation}

We can now test whether a combined variation of the two parameters $N_{eff}$ and $w$ can work jointly towards a reduction in the tension of $H_0$, i.e. for smaller values of $\Delta N_{eff}$ and $\Delta w$. By combining Eqs. \ref{novonef} and \ref{deltaw} we obtain the relation,

\begin{equation}
    \Delta H_{0}^{\Lambda CDM} = -20.6\Delta w + 5.9 \Delta N_{eff} .
    \label{delta0}
\end{equation}

Concerning possible values of $\Delta N_{eff}$, one important example is the value $\Delta N_{eff} \sim 0.11$ which is the most stringent limit  of $\Delta N_{eff}$ constrained by Planck 2018 TT,TE,EE+lowE+BAO (+He measurements) \cite{Planck:2018vyg, Aver:2015iza}.
We can analyze, as an example,  what would be the value of  $\Delta w$ necessary to solve the tension when we assume $\Delta N_{eff}=0.11$ and the Planck and SH$0$ES values for $H_0$, as

\begin{equation}
   | 73.45 - 67.72| = -20.6 \Delta w + 5.9 \times 0.11.
\end{equation}

In this case, all that is needed is a value $\Delta w \sim -0.24$. At the same time, considering another example with combined $w \sim -1.24$ and $N_{eff} \sim 3.156$, we are also able to solve the tension in the $H_0$ parameter. On the other hand, if we consider the most stringent limit from Planck 2018 \cite{Planck:2018vyg} on $w$ (i.e. $\Delta w=-0.06$, constrained at 68\% from Planck TT,TE,EE+lowE+lensing+SN+BAO data), we need a very high contribution for the number of relativistic species at recombination, of the order of $\Delta N_{eff} \sim 0.76$.

\subsection{Results on  $\Delta \sigma_{8}$}

Following the methodology used in the previous section, here we analyse the relation between the inferred values of the clustering parameter $\sigma_8$ and the variation in the values of $w$ and $N_{eff}$ in the extended models considered. 

\begin{figure}[t]
    \centering
    \includegraphics[scale=0.5]{ 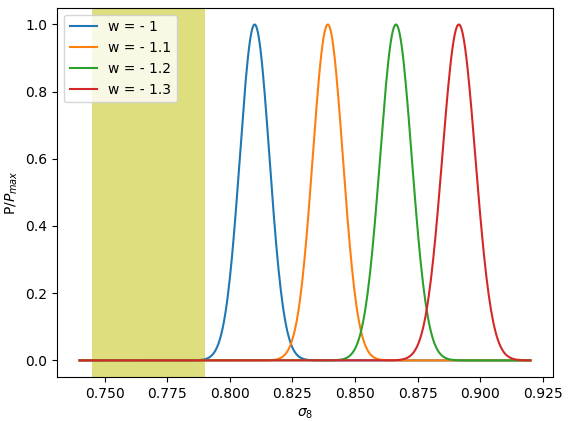}
    \caption{Normalized posterior distributions of $\sigma_8$ for different choices of w. 
    The green shaded region is the 1$\sigma$ C.L. region for $\sigma_8$ determined by KiDS-1000, yielding $\sigma_8 = 0.766^{+0.024}_{-0.021}$ \cite{KiDS:2020suj}.}
    \label{fig:sigw}
\end{figure}

Let us start again with the model $\overline{w}$CDM. In Figure \ref{fig:sigw} 
we show the normalized probability distribution function for the equation of state, $w$, compared with the Kilo-Degree Survey (KiDS-1000) lensing estimation, $\sigma_8 = 0.766^{+0.024}_{-0.021}$ \cite{KiDS:2020suj}, shown in the green shaded region.

Using the the best fit values obtained for these parameters in each model, we can relate $\Delta w$ with the inferred values for  $\Delta \sigma_{8}$ as

\begin{equation}
    \Delta \sigma_8 \equiv \sigma_8 - \sigma_8^{\Lambda CDM} \approx  -0.27(1 + w) \equiv -0.27\Delta w .
\end{equation}

As expected, the value that comes closest to the KIDS-1000 region is that for $w>-1$, which instead worsens the tension on $H_0$, as shown in Fig.\ref{fig:grafw}. This incompatibility has already been extensively discussed in the literature and makes this extension of the $\Lambda$CDM model incapable, on its own, of addressing the two tensions simultaneously.

\begin{figure}[t]
    \centering
    \includegraphics[scale=0.5]{ 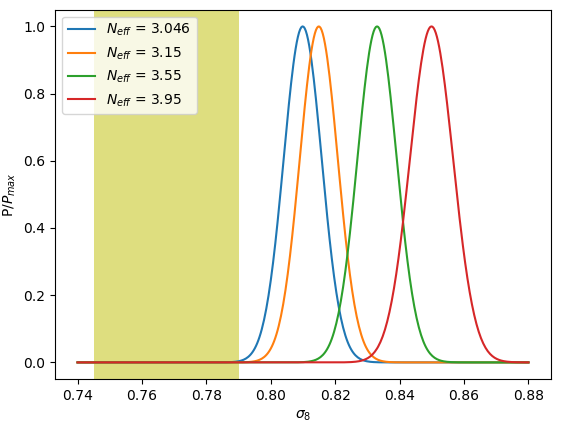}
    \caption{Normalized posterior distributions of $\sigma_8$ for different choices of $N_{eff}$. The yellow shaded region is the 1$\sigma$ C.L. region for $\sigma_8$ determined by the KiDS-1000 experiment, yielding $\sigma_8 \approx 0.766^{+0.024}_{-0.021}$ \cite{KiDS:2020suj}.}
    \label{fig:signeff}
\end{figure}

Looking at a change in the early universe using different $N_{eff}$ values in the $\overline{N}$CDM model, we see in Figure \ref{fig:signeff} that even in this case the selected values are not able to reach the KIDS-1000 region. Values lower than the standard one would allow compatibility, but at the cost of violating particle physics and considering less than three standard neutrino families.

Drawing a linear relationship using the best-fit values from our analyses we can relate  $\Delta N_{eff}$ with the inferred values for $\Delta \sigma_{8}$,

\begin{equation}
     \Delta \sigma_8 = \sigma_8 - \sigma_8^{\Lambda CDM} \approx 0.04\Delta N_{eff} = 0.04(N_{eff} - 3.046) 
\end{equation}

In this case, exploring the joint case in which we vary both $N_{eff}$ and $w$ is not of interest since, in order to relief the tension one would need to consider unphysical values of $N_{eff}<3.046$. 

\subsection{Results on $r_{s}$}

Finally, we analyse the prediction on $r_s$ for the selected models.
In this case, we choose to compare our prediction with the values of sound horizon allowed by the joint late time data of BAO and SN data, also using a prior on the $H_0$ value \cite{Bernal:2016gxb}. This region is $r_s = 136.8 \pm 4.0$.

\begin{figure}[t]
    \centering
    \includegraphics[scale=0.5]{ 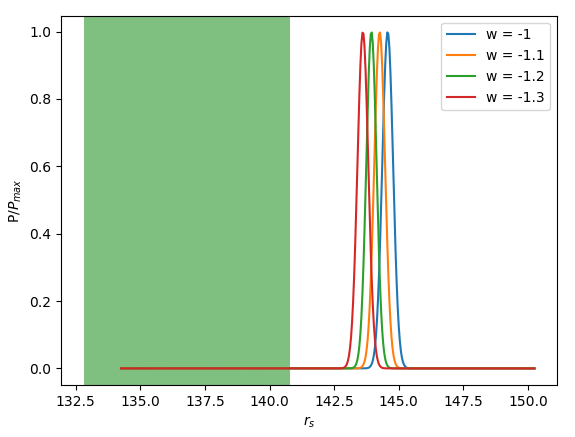}
    \caption{Normalized posterior distributions of $r_s$ for different choices of w. 
       The green shaded region is the 1$\sigma$ C.L. constrains obtained using BAO+SN+(prior on $H_0$) by Ref.
       \cite{Bernal:2016gxb}, $rs = 136.8 \pm 4.0$.}
    \label{fig:rsw}
\end{figure}

Looking at $\overline{w}$CDM case, we see in 
Figure \ref{fig:rsw} the normalized probability distribution function for different values of $w$. 
As mentioned above, varying the physics of the recent universe does not significantly affects the value of $r_s$, and in fact we note that when $w$ varies we have only a small shift in the value of the acoustic horizon at the time of recombination.  
The linear relationship between these quantities is

\begin{equation}
    \Delta r_s \equiv r_s - r_s^{\Lambda CDM}\approx 3.26 (1 + w) \equiv 3.26 \Delta w 
\end{equation}

\begin{figure}[t]
    \centering
    \includegraphics[scale=0.5]{ 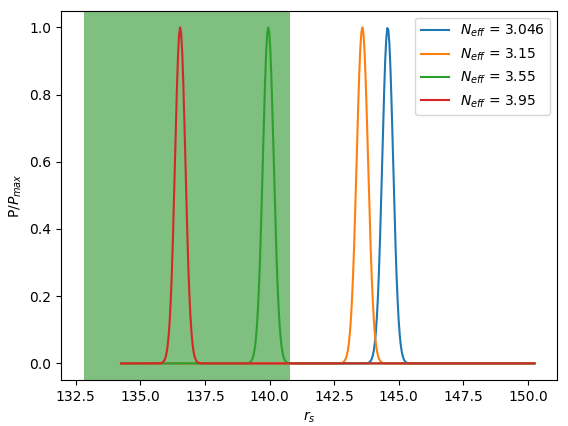}
    \caption{Normalized posterior distributions of $r_s$ for different choices of $N_{eff}$.  The green shaded region is the 1$\sigma$ C.L. constrains obtained using BAO+SN+(prior on $H_0$) by Ref.
       \cite{Bernal:2016gxb}, $rs = 136.8 \pm 4.0$.}
    \label{fig:rsneff}
\end{figure}

Instead, by introducing a modification in the primordial physics considering an extra number of relativistic species at CMB time, we can obtain a greater sensitivity in the parameter $r_s$, achieving good agreement in the case of $N_{eff} \sim 3.55$ and higher values, as showing in Figure \ref{fig:rsneff}.
 As seen in the previous sections, such a $N_{eff}$ value was able to predict an higher $H_0$ value today than that predicted by Planck (see Fig \ref{fig:grafneff}). 
The same does not apply to the $\overline{w}CDM$ model. While in Figure \ref{fig:grafw} we see that values of $w \sim 1.2 - 1.3$ allow for a prediction of $H_0$ in accordance with SH$0$ES estimations, the same is not true for retrieving values of $r_s$ in accordance with BAO+SN+$H_0$. As discussed earlier, this is because while a modification in the primordial universe affects $r_s$ but also $D_A$, the same is not true for a physics in the recent universe.

Using the best fit of our analysis we obtain 

\begin{equation}
    \Delta r_s \equiv r_s - r_s^{\Lambda CDM}\approx -8.9 (N_{eff} - 3.046) \equiv -8.9 \Delta N_{eff}
\end{equation}

\begin{figure}[t]
    \centering
    \includegraphics[scale=0.65]{ 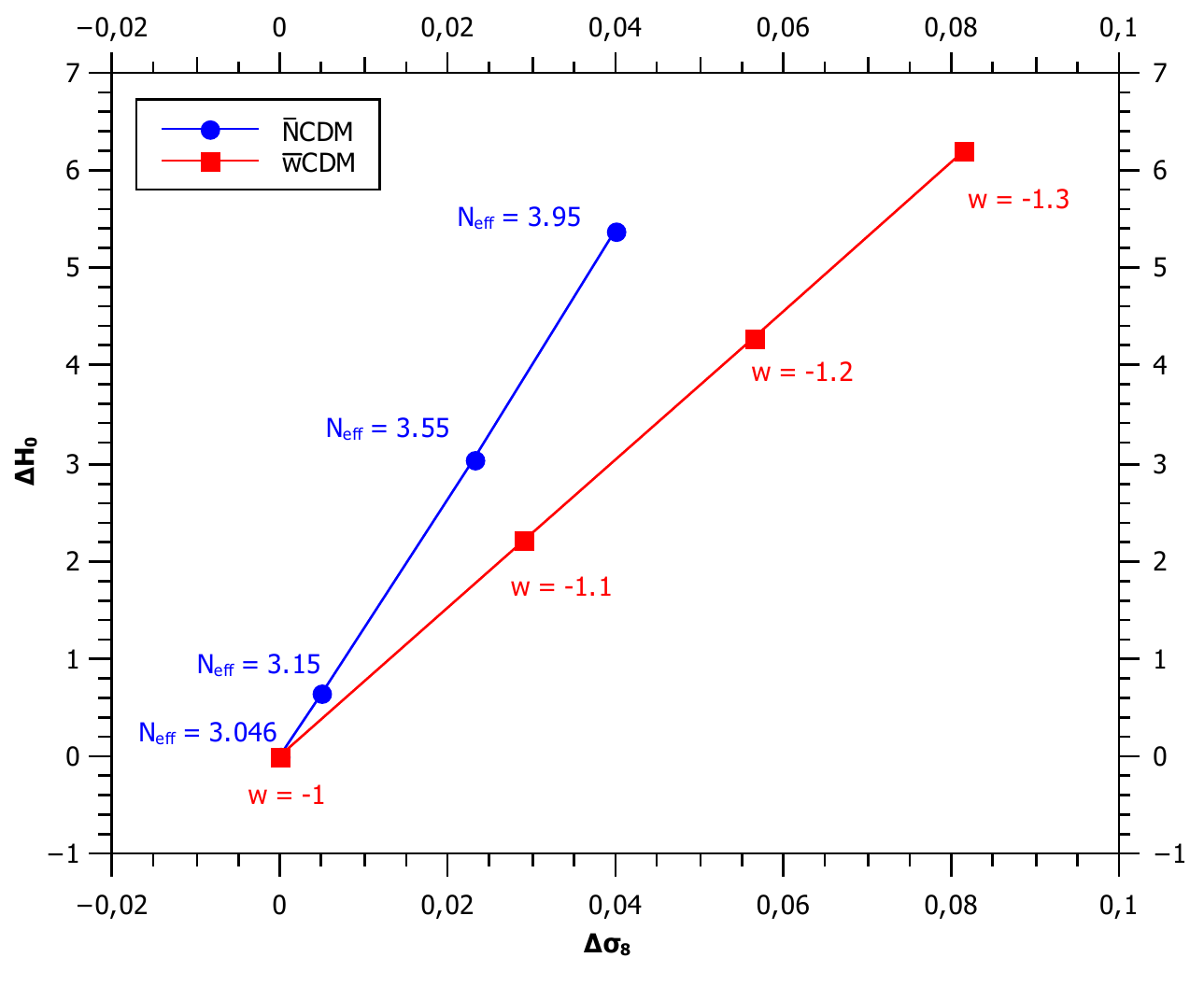}
    \caption{The $\Delta H_0$-$\Delta \sigma_8$ plane with the results of the $\overline{N}$CDM and $\overline{w}$CDM models studied.}
    \label{fig:my_label}
\end{figure}

\subsection{Relating $\Delta H_{0}$ and $\Delta \sigma_{8}$}

Following the discussion in the previous section, let us make a further step towards understanding a relationship between the variation of $H_{0}$ and $\sigma_{8}$ in the context of the extended models analyzed. 
We plot the plane $\Delta H_0$-$\Delta \sigma_8$, 
for both the $\overline{N}$CDM and $\overline{w}$CDM model, in Fig.\ref{fig:my_label}. The behaviour of the interpolations suggests that models with non-standard values for $N_{eff}$ performs better than models with non-standard values for $w$ regarding the two tensions, since for the same value of $\Delta H_0$ the first leads to smaller values of $\sigma_{8}$  than the later. This means that, compared to $\overline{N}$CDM models,  the variation in $w$ required to drive the same reduction in the $H_0$ tension  worsen the tension in  $\sigma_8$  compared to $\overline{N}$CDM models.
 One important point to take into account in comparing $\overline{N}$CDM and $\overline{w}$CDM models is how much the extra parameters are allowed to vary according to the data in each case. 
  The most stringent limit  of $\Delta N_{eff}$ in $68\%$ constrained by Planck 2018 TT,TE,EE+lowE+BAO (+He measurements) \cite{Planck:2018vyg} \cite{Aver:2015iza} corresponds to approximately $\Delta N_{eff} \sim 0.11$. 
On the other hand,  the most stringent limit from Planck 2018 \cite{Planck:2018vyg}  on $w$ corresponds to approximately $\Delta w \sim -0.06$ which is constrained at $68\%$ from Planck TT,TE,EE+lowE+lensing+SN+BAO data.

\section{Conclusions}\label{sec:Conclusions}

 The current tension   between the value inferred using early time data (i.e model dependent) and direct late time constraints (i.e model independent) on $H_{0}$ is at more than 4$\sigma$. Several possibilities have been investigated, and an exciting possibility is that this can be a sign of a new physics beyond the standard model. A tension can also be seen in the analysis of the the amplitude of the matter spectrum, parameterised by its $8$ Mpc measurement via the parameter $\sigma_8$. Indeed, a change in the value of the sound horizon (or equivantely in $H_0$) influences the inferred value of the matter density $\Omega_m$, increasing the value of the clustering parameter $\sigma_8$ and thus increasing the tension with the estimation by weak lensing data. 
 So its crucial to understand possible ways to break the correlation between these effects {
 or, also, a way of relating a dynamic measurement that comes from estimating the Hubble parameter from a model with a kinematic measurement of $H_0$, i.e. from direct observations.}
 
Usually when trying to alleviate the tensions considering models beyond the standard model it is commonly considered non-standard values for parameters which affect the constraints on $H_{0}$ and $\sigma_{8}$, 
as extra free parameters. However   marginalizing over additional parameters
implies in broadening the posterior of the cosmological
parameters correlated 
 with the additional ones. Consequently,  the Hubble tension can be relaxed  because of an increase in the
uncertainty in the $H_0$ value inferred from CMB data, and not due to a spontaneous shift in its central value. 
However, this is not the case when  a certain model predicts  a fixed non-standard value. In this case if the tension is reduced it is due to a true shift in the central value of $H_0$ or $\sigma_8$. Recently, it was analyzed the variation of $H_0$ for different fixed values of the dark energy EoS, $w$, and the $N_{eff}$ parameters, and the equation describing the linear relation between $H_{0}$ and these two parameters was obtained \cite{Vagnozzi:2019ezj}. 
 In the present work we extended these results considering a more recent data set and including also an analysis of the parameters $\sigma_{8}$ and $r_{s}$. We obtain that the parameters $H_{0}$, $\sigma_{8}$ and $r_{s}$ varies approximately linearly with the parameters $w$ and $N_{eff}$ in the interval of interest. We also obtain the adimensional multiplier describing these linear relations, providing simple expressions to quantify how the important tension parameters  vary with a change in the considered additional parameters.

Finally we also obtain a relation between $H_{0}$ and $\sigma_{8}$ for models with non-standard $w$ and non-standard $N_{eff}$. Apparently, models with non-standard values for $N_{eff}$ performs better than models with non-standard values for $w$ regarding the two tensions, since for the same value of $\Delta H_0$ the first leads to smaller values of $\Delta \sigma_{8}$  than the later. This means that, compared to $\overline{N}$CDM models,  the variation in $w$ required to drive the same reduction in the $H_0$ tension  worsen the tension in  $\sigma_{8}$  compared to $\overline{N}$CDM models.
In addition,  considering the constraints form the Planck Collaboration \cite{Planck:2018vyg} we see that the allowed limits in $\Delta w$ are more restricted than the allowed limits in $\Delta N_{eff}$.

As a future perspective, it would be interesting to further analyse the relationship between $\Delta H_0$ and $\Delta \sigma_8$ with other models that consider changes in primordial physics, to understand what characteristics a fundamental model must have in order to break the correlation between the two tensions. 
{
Although a model is still considered here to infer the value of Hubble constant and the clustering parameter, and thus falls within a \textit{dynamic estimation} of the tensions parameter, an attempt is made to minimise the variables introduced and adopt an empirical method that can give us important indications.} 
This approach is in fact a good quantitative test to check if the proposed model is working in the right direction in light of the three sensitive parameters of the debate.

\section*{Data Availability Statement}

The analysis generated during the current study are available from the corresponding author on reasonable request.

\section*{Acknowledgements}

The authors thank Elisa  G. M. Ferreira and Sunny Vagnozzi for the useful discussions.
The authors thank the use of CosmoMC code, and also acknowledge the use of the High Performance Data Center (DCON) at the Observatorio Nacional for providing the computational facilities to run our analysis. 
RS thanks support from FAPEMA and MB acknowledge Instituto Nazionale di Fisica Nucleare (INFN), Sezione di Napoli, \textit{iniziativa specifica} QGSKY.
L.L.G. would like to thank IPMU - Kavli Institute for the Physics and Mathematics of the Universe  for warm hospitality.  L.L.G is supported by CNPq, under the Grant No. 307052/2019-2, and by the Fundaçao Carlos Chagas Filho de Amparo a 
Pesquisa do Estado do Rio de Janeiro (FAPERJ), Grant No. E-26/201.297/2021.\\

\bibliography{bibliography}

\begin{thebibliography}{85}%
\makeatletter
\providecommand \@ifxundefined [1]{%
 \@ifx{#1\undefined}
}%
\providecommand \@ifnum [1]{%
 \ifnum #1\expandafter \@firstoftwo
 \else \expandafter \@secondoftwo
 \fi
}%
\providecommand \@ifx [1]{%
 \ifx #1\expandafter \@firstoftwo
 \else \expandafter \@secondoftwo
 \fi
}%
\providecommand \natexlab [1]{#1}%
\providecommand \enquote  [1]{``#1''}%
\providecommand \bibnamefont  [1]{#1}%
\providecommand \bibfnamefont [1]{#1}%
\providecommand \citenamefont [1]{#1}%
\providecommand \href@noop [0]{\@secondoftwo}%
\providecommand \href [0]{\begingroup \@sanitize@url \@href}%
\providecommand \@href[1]{\@@startlink{#1}\@@href}%
\providecommand \@@href[1]{\endgroup#1\@@endlink}%
\providecommand \@sanitize@url [0]{\catcode `\\12\catcode `\$12\catcode
  `\&12\catcode `\#12\catcode `\^12\catcode `\_12\catcode `\%12\relax}%
\providecommand \@@startlink[1]{}%
\providecommand \@@endlink[0]{}%
\providecommand \url  [0]{\begingroup\@sanitize@url \@url }%
\providecommand \@url [1]{\endgroup\@href {#1}{\urlprefix }}%
\providecommand \urlprefix  [0]{URL }%
\providecommand \Eprint [0]{\href }%
\providecommand \doibase [0]{http://dx.doi.org/}%
\providecommand \selectlanguage [0]{\@gobble}%
\providecommand \bibinfo  [0]{\@secondoftwo}%
\providecommand \bibfield  [0]{\@secondoftwo}%
\providecommand \translation [1]{[#1]}%
\providecommand \BibitemOpen [0]{}%
\providecommand \bibitemStop [0]{}%
\providecommand \bibitemNoStop [0]{.\EOS\space}%
\providecommand \EOS [0]{\spacefactor3000\relax}%
\providecommand \BibitemShut  [1]{\csname bibitem#1\endcsname}%
\let\auto@bib@innerbib\@empty
\bibitem [{\citenamefont {Aghanim}\ \emph
  {et~al.}(2020{\natexlab{a}})\citenamefont {Aghanim} \emph
  {et~al.}}]{Planck:2018vyg}%
  \BibitemOpen
  \bibfield  {author} {\bibinfo {author} {\bibfnamefont {N.}~\bibnamefont
  {Aghanim}} \emph {et~al.} (\bibinfo {collaboration} {Planck}),\ }\href
  {\doibase 10.1051/0004-6361/201833910} {\bibfield  {journal} {\bibinfo
  {journal} {Astron. Astrophys.}\ }\textbf {\bibinfo {volume} {641}},\ \bibinfo
  {pages} {A6} (\bibinfo {year} {2020}{\natexlab{a}})},\ \bibinfo {note}
  {[Erratum: Astron.Astrophys. 652, C4 (2021)]},\ \Eprint
  {http://arxiv.org/abs/1807.06209} {arXiv:1807.06209 [astro-ph.CO]}
  \BibitemShut {NoStop}%
\bibitem [{\citenamefont {Riess}\ \emph {et~al.}(2022)\citenamefont {Riess}
  \emph {et~al.}}]{Riess:2021jrx}%
  \BibitemOpen
  \bibfield  {author} {\bibinfo {author} {\bibfnamefont {A.~G.}\ \bibnamefont
  {Riess}} \emph {et~al.},\ }\href {\doibase 10.3847/2041-8213/ac5c5b}
  {\bibfield  {journal} {\bibinfo  {journal} {Astrophys. J. Lett.}\ }\textbf
  {\bibinfo {volume} {934}},\ \bibinfo {pages} {L7} (\bibinfo {year} {2022})},\
  \Eprint {http://arxiv.org/abs/2112.04510} {arXiv:2112.04510 [astro-ph.CO]}
  \BibitemShut {NoStop}%
\bibitem [{\citenamefont {Benetti}\ \emph {et~al.}(2017)\citenamefont
  {Benetti}, \citenamefont {Graef},\ and\ \citenamefont
  {Alcaniz}}]{Benetti:2017gvm}%
  \BibitemOpen
  \bibfield  {author} {\bibinfo {author} {\bibfnamefont {M.}~\bibnamefont
  {Benetti}}, \bibinfo {author} {\bibfnamefont {L.~L.}\ \bibnamefont {Graef}},
  \ and\ \bibinfo {author} {\bibfnamefont {J.~S.}\ \bibnamefont {Alcaniz}},\
  }\href {\doibase 10.1088/1475-7516/2017/04/003} {\bibfield  {journal}
  {\bibinfo  {journal} {JCAP}\ }\textbf {\bibinfo {volume} {04}},\ \bibinfo
  {pages} {003} (\bibinfo {year} {2017})},\ \Eprint
  {http://arxiv.org/abs/1702.06509} {arXiv:1702.06509 [astro-ph.CO]}
  \BibitemShut {NoStop}%
\bibitem [{\citenamefont {Benetti}\ \emph
  {et~al.}(2018{\natexlab{a}})\citenamefont {Benetti}, \citenamefont {Graef},\
  and\ \citenamefont {Alcaniz}}]{Benetti:2017juy}%
  \BibitemOpen
  \bibfield  {author} {\bibinfo {author} {\bibfnamefont {M.}~\bibnamefont
  {Benetti}}, \bibinfo {author} {\bibfnamefont {L.~L.}\ \bibnamefont {Graef}},
  \ and\ \bibinfo {author} {\bibfnamefont {J.~S.}\ \bibnamefont {Alcaniz}},\
  }\href {\doibase 10.1088/1475-7516/2018/07/066} {\bibfield  {journal}
  {\bibinfo  {journal} {JCAP}\ }\textbf {\bibinfo {volume} {07}},\ \bibinfo
  {pages} {066} (\bibinfo {year} {2018}{\natexlab{a}})},\ \Eprint
  {http://arxiv.org/abs/1712.00677} {arXiv:1712.00677 [astro-ph.CO]}
  \BibitemShut {NoStop}%
\bibitem [{\citenamefont {Vagnozzi}\ \emph
  {et~al.}(2021{\natexlab{a}})\citenamefont {Vagnozzi}, \citenamefont
  {Di~Valentino}, \citenamefont {Gariazzo}, \citenamefont {Melchiorri},
  \citenamefont {Mena},\ and\ \citenamefont {Silk}}]{Vagnozzi:2020rcz}%
  \BibitemOpen
  \bibfield  {author} {\bibinfo {author} {\bibfnamefont {S.}~\bibnamefont
  {Vagnozzi}}, \bibinfo {author} {\bibfnamefont {E.}~\bibnamefont
  {Di~Valentino}}, \bibinfo {author} {\bibfnamefont {S.}~\bibnamefont
  {Gariazzo}}, \bibinfo {author} {\bibfnamefont {A.}~\bibnamefont
  {Melchiorri}}, \bibinfo {author} {\bibfnamefont {O.}~\bibnamefont {Mena}}, \
  and\ \bibinfo {author} {\bibfnamefont {J.}~\bibnamefont {Silk}},\ }\href
  {\doibase 10.1016/j.dark.2021.100851} {\bibfield  {journal} {\bibinfo
  {journal} {Phys. Dark Univ.}\ }\textbf {\bibinfo {volume} {33}},\ \bibinfo
  {pages} {100851} (\bibinfo {year} {2021}{\natexlab{a}})},\ \Eprint
  {http://arxiv.org/abs/2010.02230} {arXiv:2010.02230 [astro-ph.CO]}
  \BibitemShut {NoStop}%
\bibitem [{\citenamefont {Di~Valentino}\ \emph
  {et~al.}(2021{\natexlab{a}})\citenamefont {Di~Valentino}, \citenamefont
  {Melchiorri},\ and\ \citenamefont {Silk}}]{DiValentino:2020hov}%
  \BibitemOpen
  \bibfield  {author} {\bibinfo {author} {\bibfnamefont {E.}~\bibnamefont
  {Di~Valentino}}, \bibinfo {author} {\bibfnamefont {A.}~\bibnamefont
  {Melchiorri}}, \ and\ \bibinfo {author} {\bibfnamefont {J.}~\bibnamefont
  {Silk}},\ }\href {\doibase 10.3847/2041-8213/abe1c4} {\bibfield  {journal}
  {\bibinfo  {journal} {Astrophys. J. Lett.}\ }\textbf {\bibinfo {volume}
  {908}},\ \bibinfo {pages} {L9} (\bibinfo {year} {2021}{\natexlab{a}})},\
  \Eprint {http://arxiv.org/abs/2003.04935} {arXiv:2003.04935 [astro-ph.CO]}
  \BibitemShut {NoStop}%
\bibitem [{\citenamefont {Bose}\ and\ \citenamefont
  {Lombriser}(2021)}]{Bose:2020cjb}%
  \BibitemOpen
  \bibfield  {author} {\bibinfo {author} {\bibfnamefont {B.}~\bibnamefont
  {Bose}}\ and\ \bibinfo {author} {\bibfnamefont {L.}~\bibnamefont
  {Lombriser}},\ }\href {\doibase 10.1103/PhysRevD.103.L081304} {\bibfield
  {journal} {\bibinfo  {journal} {Phys. Rev. D}\ }\textbf {\bibinfo {volume}
  {103}},\ \bibinfo {pages} {L081304} (\bibinfo {year} {2021})},\ \Eprint
  {http://arxiv.org/abs/2006.16149} {arXiv:2006.16149 [astro-ph.CO]}
  \BibitemShut {NoStop}%
\bibitem [{\citenamefont {Gonzalez}\ \emph {et~al.}(2021)\citenamefont
  {Gonzalez}, \citenamefont {Benetti}, \citenamefont {von Marttens},\ and\
  \citenamefont {Alcaniz}}]{Gonzalez:2021ojp}%
  \BibitemOpen
  \bibfield  {author} {\bibinfo {author} {\bibfnamefont {J.~E.}\ \bibnamefont
  {Gonzalez}}, \bibinfo {author} {\bibfnamefont {M.}~\bibnamefont {Benetti}},
  \bibinfo {author} {\bibfnamefont {R.}~\bibnamefont {von Marttens}}, \ and\
  \bibinfo {author} {\bibfnamefont {J.}~\bibnamefont {Alcaniz}},\ }\href
  {\doibase 10.1088/1475-7516/2021/11/060} {\bibfield  {journal} {\bibinfo
  {journal} {JCAP}\ }\textbf {\bibinfo {volume} {11}},\ \bibinfo {pages} {060}
  (\bibinfo {year} {2021})},\ \Eprint {http://arxiv.org/abs/2104.13455}
  {arXiv:2104.13455 [astro-ph.CO]} \BibitemShut {NoStop}%
\bibitem [{\citenamefont {Efstathiou}\ and\ \citenamefont
  {Gratton}(2020)}]{Efstathiou:2020wem}%
  \BibitemOpen
  \bibfield  {author} {\bibinfo {author} {\bibfnamefont {G.}~\bibnamefont
  {Efstathiou}}\ and\ \bibinfo {author} {\bibfnamefont {S.}~\bibnamefont
  {Gratton}},\ }\href {\doibase 10.1093/mnrasl/slaa093} {\bibfield  {journal}
  {\bibinfo  {journal} {Mon. Not. Roy. Astron. Soc.}\ }\textbf {\bibinfo
  {volume} {496}},\ \bibinfo {pages} {L91} (\bibinfo {year} {2020})},\ \Eprint
  {http://arxiv.org/abs/2002.06892} {arXiv:2002.06892 [astro-ph.CO]}
  \BibitemShut {NoStop}%
\bibitem [{\citenamefont {Vagnozzi}\ \emph
  {et~al.}(2021{\natexlab{b}})\citenamefont {Vagnozzi}, \citenamefont {Loeb},\
  and\ \citenamefont {Moresco}}]{Vagnozzi:2020dfn}%
  \BibitemOpen
  \bibfield  {author} {\bibinfo {author} {\bibfnamefont {S.}~\bibnamefont
  {Vagnozzi}}, \bibinfo {author} {\bibfnamefont {A.}~\bibnamefont {Loeb}}, \
  and\ \bibinfo {author} {\bibfnamefont {M.}~\bibnamefont {Moresco}},\ }\href
  {\doibase 10.3847/1538-4357/abd4df} {\bibfield  {journal} {\bibinfo
  {journal} {Astrophys. J.}\ }\textbf {\bibinfo {volume} {908}},\ \bibinfo
  {pages} {84} (\bibinfo {year} {2021}{\natexlab{b}})},\ \Eprint
  {http://arxiv.org/abs/2011.11645} {arXiv:2011.11645 [astro-ph.CO]}
  \BibitemShut {NoStop}%
\bibitem [{\citenamefont {Ivanov}\ \emph {et~al.}(2020)\citenamefont {Ivanov},
  \citenamefont {Ali-Ha\"\i{}moud},\ and\ \citenamefont
  {Lesgourgues}}]{Ivanov:2020mfr}%
  \BibitemOpen
  \bibfield  {author} {\bibinfo {author} {\bibfnamefont {M.~M.}\ \bibnamefont
  {Ivanov}}, \bibinfo {author} {\bibfnamefont {Y.}~\bibnamefont
  {Ali-Ha\"\i{}moud}}, \ and\ \bibinfo {author} {\bibfnamefont
  {J.}~\bibnamefont {Lesgourgues}},\ }\href {\doibase
  10.1103/PhysRevD.102.063515} {\bibfield  {journal} {\bibinfo  {journal}
  {Phys. Rev. D}\ }\textbf {\bibinfo {volume} {102}},\ \bibinfo {pages}
  {063515} (\bibinfo {year} {2020})},\ \Eprint
  {http://arxiv.org/abs/2005.10656} {arXiv:2005.10656 [astro-ph.CO]}
  \BibitemShut {NoStop}%
\bibitem [{\citenamefont {Di~Valentino}\ \emph
  {et~al.}(2021{\natexlab{b}})\citenamefont {Di~Valentino}, \citenamefont
  {Mena}, \citenamefont {Pan}, \citenamefont {Visinelli}, \citenamefont {Yang},
  \citenamefont {Melchiorri}, \citenamefont {Mota}, \citenamefont {Riess},\
  and\ \citenamefont {Silk}}]{DiValentino:2021izs}%
  \BibitemOpen
  \bibfield  {author} {\bibinfo {author} {\bibfnamefont {E.}~\bibnamefont
  {Di~Valentino}}, \bibinfo {author} {\bibfnamefont {O.}~\bibnamefont {Mena}},
  \bibinfo {author} {\bibfnamefont {S.}~\bibnamefont {Pan}}, \bibinfo {author}
  {\bibfnamefont {L.}~\bibnamefont {Visinelli}}, \bibinfo {author}
  {\bibfnamefont {W.}~\bibnamefont {Yang}}, \bibinfo {author} {\bibfnamefont
  {A.}~\bibnamefont {Melchiorri}}, \bibinfo {author} {\bibfnamefont {D.~F.}\
  \bibnamefont {Mota}}, \bibinfo {author} {\bibfnamefont {A.~G.}\ \bibnamefont
  {Riess}}, \ and\ \bibinfo {author} {\bibfnamefont {J.}~\bibnamefont {Silk}},\
  }\href {\doibase 10.1088/1361-6382/ac086d} {\bibfield  {journal} {\bibinfo
  {journal} {Class. Quant. Grav.}\ }\textbf {\bibinfo {volume} {38}},\ \bibinfo
  {pages} {153001} (\bibinfo {year} {2021}{\natexlab{b}})},\ \Eprint
  {http://arxiv.org/abs/2103.01183} {arXiv:2103.01183 [astro-ph.CO]}
  \BibitemShut {NoStop}%
\bibitem [{\citenamefont {Herold}\ \emph {et~al.}(2022)\citenamefont {Herold},
  \citenamefont {Ferreira},\ and\ \citenamefont {Komatsu}}]{Herold:2021ksg}%
  \BibitemOpen
  \bibfield  {author} {\bibinfo {author} {\bibfnamefont {L.}~\bibnamefont
  {Herold}}, \bibinfo {author} {\bibfnamefont {E.~G.~M.}\ \bibnamefont
  {Ferreira}}, \ and\ \bibinfo {author} {\bibfnamefont {E.}~\bibnamefont
  {Komatsu}},\ }\href {\doibase 10.3847/2041-8213/ac63a3} {\bibfield  {journal}
  {\bibinfo  {journal} {Astrophys. J. Lett.}\ }\textbf {\bibinfo {volume}
  {929}},\ \bibinfo {pages} {L16} (\bibinfo {year} {2022})},\ \Eprint
  {http://arxiv.org/abs/2112.12140} {arXiv:2112.12140 [astro-ph.CO]}
  \BibitemShut {NoStop}%
\bibitem [{\citenamefont {Reeves}\ \emph {et~al.}(2022)\citenamefont {Reeves},
  \citenamefont {Herold}, \citenamefont {Vagnozzi}, \citenamefont {Sherwin},\
  and\ \citenamefont {Ferreira}}]{Reeves:2022aoi}%
  \BibitemOpen
  \bibfield  {author} {\bibinfo {author} {\bibfnamefont {A.}~\bibnamefont
  {Reeves}}, \bibinfo {author} {\bibfnamefont {L.}~\bibnamefont {Herold}},
  \bibinfo {author} {\bibfnamefont {S.}~\bibnamefont {Vagnozzi}}, \bibinfo
  {author} {\bibfnamefont {B.~D.}\ \bibnamefont {Sherwin}}, \ and\ \bibinfo
  {author} {\bibfnamefont {E.~G.~M.}\ \bibnamefont {Ferreira}},\ }\href@noop {}
  {\  (\bibinfo {year} {2022})},\ \Eprint {http://arxiv.org/abs/2207.01501}
  {arXiv:2207.01501 [astro-ph.CO]} \BibitemShut {NoStop}%
\bibitem [{\citenamefont {Karwal}\ and\ \citenamefont
  {Kamionkowski}(2016)}]{Karwal:2016vyq}%
  \BibitemOpen
  \bibfield  {author} {\bibinfo {author} {\bibfnamefont {T.}~\bibnamefont
  {Karwal}}\ and\ \bibinfo {author} {\bibfnamefont {M.}~\bibnamefont
  {Kamionkowski}},\ }\href {\doibase 10.1103/PhysRevD.94.103523} {\bibfield
  {journal} {\bibinfo  {journal} {Phys. Rev. D}\ }\textbf {\bibinfo {volume}
  {94}},\ \bibinfo {pages} {103523} (\bibinfo {year} {2016})},\ \Eprint
  {http://arxiv.org/abs/1608.01309} {arXiv:1608.01309 [astro-ph.CO]}
  \BibitemShut {NoStop}%
\bibitem [{\citenamefont {Vagnozzi}(2021{\natexlab{a}})}]{Vagnozzi:2021gjh}%
  \BibitemOpen
  \bibfield  {author} {\bibinfo {author} {\bibfnamefont {S.}~\bibnamefont
  {Vagnozzi}},\ }\href {\doibase 10.1103/PhysRevD.104.063524} {\bibfield
  {journal} {\bibinfo  {journal} {Phys. Rev. D}\ }\textbf {\bibinfo {volume}
  {104}},\ \bibinfo {pages} {063524} (\bibinfo {year} {2021}{\natexlab{a}})},\
  \Eprint {http://arxiv.org/abs/2105.10425} {arXiv:2105.10425 [astro-ph.CO]}
  \BibitemShut {NoStop}%
\bibitem [{\citenamefont {Zhao}\ \emph {et~al.}(2017)\citenamefont {Zhao} \emph
  {et~al.}}]{Zhao:2017cud}%
  \BibitemOpen
  \bibfield  {author} {\bibinfo {author} {\bibfnamefont {G.-B.}\ \bibnamefont
  {Zhao}} \emph {et~al.},\ }\href {\doibase 10.1038/s41550-017-0216-z}
  {\bibfield  {journal} {\bibinfo  {journal} {Nature Astron.}\ }\textbf
  {\bibinfo {volume} {1}},\ \bibinfo {pages} {627} (\bibinfo {year} {2017})},\
  \Eprint {http://arxiv.org/abs/1701.08165} {arXiv:1701.08165 [astro-ph.CO]}
  \BibitemShut {NoStop}%
\bibitem [{\citenamefont {Bolotin}\ \emph {et~al.}(2014)\citenamefont
  {Bolotin}, \citenamefont {Kostenko}, \citenamefont {Lemets},\ and\
  \citenamefont {Yerokhin}}]{Bolotin:2013jpa}%
  \BibitemOpen
  \bibfield  {author} {\bibinfo {author} {\bibfnamefont {Y.~L.}\ \bibnamefont
  {Bolotin}}, \bibinfo {author} {\bibfnamefont {A.}~\bibnamefont {Kostenko}},
  \bibinfo {author} {\bibfnamefont {O.~A.}\ \bibnamefont {Lemets}}, \ and\
  \bibinfo {author} {\bibfnamefont {D.~A.}\ \bibnamefont {Yerokhin}},\ }\href
  {\doibase 10.1142/S0218271815300074} {\bibfield  {journal} {\bibinfo
  {journal} {Int. J. Mod. Phys. D}\ }\textbf {\bibinfo {volume} {24}},\
  \bibinfo {pages} {1530007} (\bibinfo {year} {2014})},\ \Eprint
  {http://arxiv.org/abs/1310.0085} {arXiv:1310.0085 [astro-ph.CO]} \BibitemShut
  {NoStop}%
\bibitem [{\citenamefont {Wang}\ \emph {et~al.}(2016)\citenamefont {Wang},
  \citenamefont {Abdalla}, \citenamefont {Atrio-Barandela},\ and\ \citenamefont
  {Pavon}}]{Wang:2016lxa}%
  \BibitemOpen
  \bibfield  {author} {\bibinfo {author} {\bibfnamefont {B.}~\bibnamefont
  {Wang}}, \bibinfo {author} {\bibfnamefont {E.}~\bibnamefont {Abdalla}},
  \bibinfo {author} {\bibfnamefont {F.}~\bibnamefont {Atrio-Barandela}}, \ and\
  \bibinfo {author} {\bibfnamefont {D.}~\bibnamefont {Pavon}},\ }\href
  {\doibase 10.1088/0034-4885/79/9/096901} {\bibfield  {journal} {\bibinfo
  {journal} {Rept. Prog. Phys.}\ }\textbf {\bibinfo {volume} {79}},\ \bibinfo
  {pages} {096901} (\bibinfo {year} {2016})},\ \Eprint
  {http://arxiv.org/abs/1603.08299} {arXiv:1603.08299 [astro-ph.CO]}
  \BibitemShut {NoStop}%
\bibitem [{\citenamefont {von Marttens}\ \emph {et~al.}(2021)\citenamefont {von
  Marttens}, \citenamefont {Gonzalez}, \citenamefont {Alcaniz}, \citenamefont
  {Marra},\ and\ \citenamefont {Casarini}}]{vonMarttens:2020apn}%
  \BibitemOpen
  \bibfield  {author} {\bibinfo {author} {\bibfnamefont {R.}~\bibnamefont {von
  Marttens}}, \bibinfo {author} {\bibfnamefont {J.~E.}\ \bibnamefont
  {Gonzalez}}, \bibinfo {author} {\bibfnamefont {J.}~\bibnamefont {Alcaniz}},
  \bibinfo {author} {\bibfnamefont {V.}~\bibnamefont {Marra}}, \ and\ \bibinfo
  {author} {\bibfnamefont {L.}~\bibnamefont {Casarini}},\ }\href {\doibase
  10.1103/PhysRevD.104.043515} {\bibfield  {journal} {\bibinfo  {journal}
  {Phys. Rev. D}\ }\textbf {\bibinfo {volume} {104}},\ \bibinfo {pages}
  {043515} (\bibinfo {year} {2021})},\ \Eprint
  {http://arxiv.org/abs/2011.10846} {arXiv:2011.10846 [astro-ph.CO]}
  \BibitemShut {NoStop}%
\bibitem [{\citenamefont {von Marttens}\ \emph {et~al.}(2022)\citenamefont {von
  Marttens}, \citenamefont {Barbosa},\ and\ \citenamefont
  {Alcaniz}}]{vonMarttens:2022xyr}%
  \BibitemOpen
  \bibfield  {author} {\bibinfo {author} {\bibfnamefont {R.}~\bibnamefont {von
  Marttens}}, \bibinfo {author} {\bibfnamefont {D.}~\bibnamefont {Barbosa}}, \
  and\ \bibinfo {author} {\bibfnamefont {J.}~\bibnamefont {Alcaniz}},\
  }\href@noop {} {\  (\bibinfo {year} {2022})},\ \Eprint
  {http://arxiv.org/abs/2208.06302} {arXiv:2208.06302 [astro-ph.CO]}
  \BibitemShut {NoStop}%
\bibitem [{\citenamefont {Benetti}\ \emph {et~al.}(2019)\citenamefont
  {Benetti}, \citenamefont {Miranda}, \citenamefont {Borges}, \citenamefont
  {Pigozzo}, \citenamefont {Carneiro},\ and\ \citenamefont
  {Alcaniz}}]{Benetti:2019lxu}%
  \BibitemOpen
  \bibfield  {author} {\bibinfo {author} {\bibfnamefont {M.}~\bibnamefont
  {Benetti}}, \bibinfo {author} {\bibfnamefont {W.}~\bibnamefont {Miranda}},
  \bibinfo {author} {\bibfnamefont {H.~A.}\ \bibnamefont {Borges}}, \bibinfo
  {author} {\bibfnamefont {C.}~\bibnamefont {Pigozzo}}, \bibinfo {author}
  {\bibfnamefont {S.}~\bibnamefont {Carneiro}}, \ and\ \bibinfo {author}
  {\bibfnamefont {J.~S.}\ \bibnamefont {Alcaniz}},\ }\href {\doibase
  10.1088/1475-7516/2019/12/023} {\bibfield  {journal} {\bibinfo  {journal}
  {JCAP}\ }\textbf {\bibinfo {volume} {12}},\ \bibinfo {pages} {023} (\bibinfo
  {year} {2019})},\ \Eprint {http://arxiv.org/abs/1908.07213} {arXiv:1908.07213
  [astro-ph.CO]} \BibitemShut {NoStop}%
\bibitem [{\citenamefont {Salzano}\ \emph {et~al.}(2021)\citenamefont {Salzano}
  \emph {et~al.}}]{Salzano:2021zxk}%
  \BibitemOpen
  \bibfield  {author} {\bibinfo {author} {\bibfnamefont {V.}~\bibnamefont
  {Salzano}} \emph {et~al.},\ }\href {\doibase 10.1088/1475-7516/2021/09/033}
  {\bibfield  {journal} {\bibinfo  {journal} {JCAP}\ }\textbf {\bibinfo
  {volume} {09}},\ \bibinfo {pages} {033} (\bibinfo {year} {2021})},\ \Eprint
  {http://arxiv.org/abs/2102.06417} {arXiv:2102.06417 [astro-ph.CO]}
  \BibitemShut {NoStop}%
\bibitem [{\citenamefont {Benetti}\ \emph {et~al.}(2021)\citenamefont
  {Benetti}, \citenamefont {Borges}, \citenamefont {Pigozzo}, \citenamefont
  {Carneiro},\ and\ \citenamefont {Alcaniz}}]{Benetti:2021div}%
  \BibitemOpen
  \bibfield  {author} {\bibinfo {author} {\bibfnamefont {M.}~\bibnamefont
  {Benetti}}, \bibinfo {author} {\bibfnamefont {H.}~\bibnamefont {Borges}},
  \bibinfo {author} {\bibfnamefont {C.}~\bibnamefont {Pigozzo}}, \bibinfo
  {author} {\bibfnamefont {S.}~\bibnamefont {Carneiro}}, \ and\ \bibinfo
  {author} {\bibfnamefont {J.}~\bibnamefont {Alcaniz}},\ }\href {\doibase
  10.1088/1475-7516/2021/08/014} {\bibfield  {journal} {\bibinfo  {journal}
  {JCAP}\ }\textbf {\bibinfo {volume} {08}},\ \bibinfo {pages} {014} (\bibinfo
  {year} {2021})},\ \Eprint {http://arxiv.org/abs/2102.10123} {arXiv:2102.10123
  [astro-ph.CO]} \BibitemShut {NoStop}%
\bibitem [{\citenamefont {Di~Valentino}\ \emph
  {et~al.}(2020{\natexlab{a}})\citenamefont {Di~Valentino}, \citenamefont
  {Melchiorri}, \citenamefont {Mena},\ and\ \citenamefont
  {Vagnozzi}}]{DiValentino:2019ffd}%
  \BibitemOpen
  \bibfield  {author} {\bibinfo {author} {\bibfnamefont {E.}~\bibnamefont
  {Di~Valentino}}, \bibinfo {author} {\bibfnamefont {A.}~\bibnamefont
  {Melchiorri}}, \bibinfo {author} {\bibfnamefont {O.}~\bibnamefont {Mena}}, \
  and\ \bibinfo {author} {\bibfnamefont {S.}~\bibnamefont {Vagnozzi}},\ }\href
  {\doibase 10.1016/j.dark.2020.100666} {\bibfield  {journal} {\bibinfo
  {journal} {Phys. Dark Univ.}\ }\textbf {\bibinfo {volume} {30}},\ \bibinfo
  {pages} {100666} (\bibinfo {year} {2020}{\natexlab{a}})},\ \Eprint
  {http://arxiv.org/abs/1908.04281} {arXiv:1908.04281 [astro-ph.CO]}
  \BibitemShut {NoStop}%
\bibitem [{\citenamefont {Kuroyanagi}\ \emph {et~al.}(2021)\citenamefont
  {Kuroyanagi}, \citenamefont {Takahashi},\ and\ \citenamefont
  {Yokoyama}}]{Kuroyanagi:2020sfw}%
  \BibitemOpen
  \bibfield  {author} {\bibinfo {author} {\bibfnamefont {S.}~\bibnamefont
  {Kuroyanagi}}, \bibinfo {author} {\bibfnamefont {T.}~\bibnamefont
  {Takahashi}}, \ and\ \bibinfo {author} {\bibfnamefont {S.}~\bibnamefont
  {Yokoyama}},\ }\href {\doibase 10.1088/1475-7516/2021/01/071} {\bibfield
  {journal} {\bibinfo  {journal} {JCAP}\ }\textbf {\bibinfo {volume} {01}},\
  \bibinfo {pages} {071} (\bibinfo {year} {2021})},\ \Eprint
  {http://arxiv.org/abs/2011.03323} {arXiv:2011.03323 [astro-ph.CO]}
  \BibitemShut {NoStop}%
\bibitem [{\citenamefont {Dom\`enech}\ \emph {et~al.}(2020)\citenamefont
  {Dom\`enech}, \citenamefont {Pi},\ and\ \citenamefont
  {Sasaki}}]{Domenech:2020kqm}%
  \BibitemOpen
  \bibfield  {author} {\bibinfo {author} {\bibfnamefont {G.}~\bibnamefont
  {Dom\`enech}}, \bibinfo {author} {\bibfnamefont {S.}~\bibnamefont {Pi}}, \
  and\ \bibinfo {author} {\bibfnamefont {M.}~\bibnamefont {Sasaki}},\ }\href
  {\doibase 10.1088/1475-7516/2020/08/017} {\bibfield  {journal} {\bibinfo
  {journal} {JCAP}\ }\textbf {\bibinfo {volume} {08}},\ \bibinfo {pages} {017}
  (\bibinfo {year} {2020})},\ \Eprint {http://arxiv.org/abs/2005.12314}
  {arXiv:2005.12314 [gr-qc]} \BibitemShut {NoStop}%
\bibitem [{\citenamefont {Vagnozzi}(2021{\natexlab{b}})}]{Vagnozzi:2020gtf}%
  \BibitemOpen
  \bibfield  {author} {\bibinfo {author} {\bibfnamefont {S.}~\bibnamefont
  {Vagnozzi}},\ }\href {\doibase 10.1093/mnrasl/slaa203} {\bibfield  {journal}
  {\bibinfo  {journal} {Mon. Not. Roy. Astron. Soc.}\ }\textbf {\bibinfo
  {volume} {502}},\ \bibinfo {pages} {L11} (\bibinfo {year}
  {2021}{\natexlab{b}})},\ \Eprint {http://arxiv.org/abs/2009.13432}
  {arXiv:2009.13432 [astro-ph.CO]} \BibitemShut {NoStop}%
\bibitem [{\citenamefont {Sugiyama}\ \emph {et~al.}(2021)\citenamefont
  {Sugiyama}, \citenamefont {Takhistov}, \citenamefont {Vitagliano},
  \citenamefont {Kusenko}, \citenamefont {Sasaki},\ and\ \citenamefont
  {Takada}}]{Sugiyama:2020roc}%
  \BibitemOpen
  \bibfield  {author} {\bibinfo {author} {\bibfnamefont {S.}~\bibnamefont
  {Sugiyama}}, \bibinfo {author} {\bibfnamefont {V.}~\bibnamefont {Takhistov}},
  \bibinfo {author} {\bibfnamefont {E.}~\bibnamefont {Vitagliano}}, \bibinfo
  {author} {\bibfnamefont {A.}~\bibnamefont {Kusenko}}, \bibinfo {author}
  {\bibfnamefont {M.}~\bibnamefont {Sasaki}}, \ and\ \bibinfo {author}
  {\bibfnamefont {M.}~\bibnamefont {Takada}},\ }\href {\doibase
  10.1016/j.physletb.2021.136097} {\bibfield  {journal} {\bibinfo  {journal}
  {Phys. Lett. B}\ }\textbf {\bibinfo {volume} {814}},\ \bibinfo {pages}
  {136097} (\bibinfo {year} {2021})},\ \Eprint
  {http://arxiv.org/abs/2010.02189} {arXiv:2010.02189 [astro-ph.CO]}
  \BibitemShut {NoStop}%
\bibitem [{\citenamefont {Namba}\ and\ \citenamefont
  {Suzuki}(2020)}]{Namba:2020kij}%
  \BibitemOpen
  \bibfield  {author} {\bibinfo {author} {\bibfnamefont {R.}~\bibnamefont
  {Namba}}\ and\ \bibinfo {author} {\bibfnamefont {M.}~\bibnamefont {Suzuki}},\
  }\href {\doibase 10.1103/PhysRevD.102.123527} {\bibfield  {journal} {\bibinfo
   {journal} {Phys. Rev. D}\ }\textbf {\bibinfo {volume} {102}},\ \bibinfo
  {pages} {123527} (\bibinfo {year} {2020})},\ \Eprint
  {http://arxiv.org/abs/2009.13909} {arXiv:2009.13909 [astro-ph.CO]}
  \BibitemShut {NoStop}%
\bibitem [{\citenamefont {Graef}\ \emph {et~al.}(2019)\citenamefont {Graef},
  \citenamefont {Benetti},\ and\ \citenamefont {Alcaniz}}]{Graef:2018fzu}%
  \BibitemOpen
  \bibfield  {author} {\bibinfo {author} {\bibfnamefont {L.~L.}\ \bibnamefont
  {Graef}}, \bibinfo {author} {\bibfnamefont {M.}~\bibnamefont {Benetti}}, \
  and\ \bibinfo {author} {\bibfnamefont {J.~S.}\ \bibnamefont {Alcaniz}},\
  }\href {\doibase 10.1103/PhysRevD.99.043519} {\bibfield  {journal} {\bibinfo
  {journal} {Phys. Rev. D}\ }\textbf {\bibinfo {volume} {99}},\ \bibinfo
  {pages} {043519} (\bibinfo {year} {2019})},\ \Eprint
  {http://arxiv.org/abs/1809.04501} {arXiv:1809.04501 [astro-ph.CO]}
  \BibitemShut {NoStop}%
\bibitem [{\citenamefont {Kuroyanagi}\ \emph {et~al.}(2015)\citenamefont
  {Kuroyanagi}, \citenamefont {Takahashi},\ and\ \citenamefont
  {Yokoyama}}]{Kuroyanagi:2014nba}%
  \BibitemOpen
  \bibfield  {author} {\bibinfo {author} {\bibfnamefont {S.}~\bibnamefont
  {Kuroyanagi}}, \bibinfo {author} {\bibfnamefont {T.}~\bibnamefont
  {Takahashi}}, \ and\ \bibinfo {author} {\bibfnamefont {S.}~\bibnamefont
  {Yokoyama}},\ }\href {\doibase 10.1088/1475-7516/2015/02/003} {\bibfield
  {journal} {\bibinfo  {journal} {JCAP}\ }\textbf {\bibinfo {volume} {02}},\
  \bibinfo {pages} {003} (\bibinfo {year} {2015})},\ \Eprint
  {http://arxiv.org/abs/1407.4785} {arXiv:1407.4785 [astro-ph.CO]} \BibitemShut
  {NoStop}%
\bibitem [{\citenamefont {Aloni}\ \emph {et~al.}(2022)\citenamefont {Aloni},
  \citenamefont {Berlin}, \citenamefont {Joseph}, \citenamefont {Schmaltz},\
  and\ \citenamefont {Weiner}}]{Aloni:2021eaq}%
  \BibitemOpen
  \bibfield  {author} {\bibinfo {author} {\bibfnamefont {D.}~\bibnamefont
  {Aloni}}, \bibinfo {author} {\bibfnamefont {A.}~\bibnamefont {Berlin}},
  \bibinfo {author} {\bibfnamefont {M.}~\bibnamefont {Joseph}}, \bibinfo
  {author} {\bibfnamefont {M.}~\bibnamefont {Schmaltz}}, \ and\ \bibinfo
  {author} {\bibfnamefont {N.}~\bibnamefont {Weiner}},\ }\href {\doibase
  10.1103/PhysRevD.105.123516} {\bibfield  {journal} {\bibinfo  {journal}
  {Phys. Rev. D}\ }\textbf {\bibinfo {volume} {105}},\ \bibinfo {pages}
  {123516} (\bibinfo {year} {2022})},\ \Eprint
  {http://arxiv.org/abs/2111.00014} {arXiv:2111.00014 [astro-ph.CO]}
  \BibitemShut {NoStop}%
\bibitem [{\citenamefont {Joseph}\ \emph {et~al.}(2022)\citenamefont {Joseph},
  \citenamefont {Aloni}, \citenamefont {Schmaltz}, \citenamefont {Sivarajan},\
  and\ \citenamefont {Weiner}}]{Joseph:2022jsf}%
  \BibitemOpen
  \bibfield  {author} {\bibinfo {author} {\bibfnamefont {M.}~\bibnamefont
  {Joseph}}, \bibinfo {author} {\bibfnamefont {D.}~\bibnamefont {Aloni}},
  \bibinfo {author} {\bibfnamefont {M.}~\bibnamefont {Schmaltz}}, \bibinfo
  {author} {\bibfnamefont {E.~N.}\ \bibnamefont {Sivarajan}}, \ and\ \bibinfo
  {author} {\bibfnamefont {N.}~\bibnamefont {Weiner}},\ }\href@noop {} {\
  (\bibinfo {year} {2022})},\ \Eprint {http://arxiv.org/abs/2207.03500}
  {arXiv:2207.03500 [astro-ph.CO]} \BibitemShut {NoStop}%
\bibitem [{\citenamefont {Sch\"oneberg}\ and\ \citenamefont
  {Franco~Abell\'an}(2022)}]{Schoneberg:2022grr}%
  \BibitemOpen
  \bibfield  {author} {\bibinfo {author} {\bibfnamefont {N.}~\bibnamefont
  {Sch\"oneberg}}\ and\ \bibinfo {author} {\bibfnamefont {G.}~\bibnamefont
  {Franco~Abell\'an}},\ }\href@noop {} {\  (\bibinfo {year} {2022})},\ \Eprint
  {http://arxiv.org/abs/2206.11276} {arXiv:2206.11276 [astro-ph.CO]}
  \BibitemShut {NoStop}%
\bibitem [{\citenamefont {Hova}\ and\ \citenamefont
  {Yang}(2017)}]{Hova:2010na}%
  \BibitemOpen
  \bibfield  {author} {\bibinfo {author} {\bibfnamefont {H.}~\bibnamefont
  {Hova}}\ and\ \bibinfo {author} {\bibfnamefont {H.}~\bibnamefont {Yang}},\
  }\href {\doibase 10.1142/S0218271817501784} {\bibfield  {journal} {\bibinfo
  {journal} {Int. J. Mod. Phys. D}\ }\textbf {\bibinfo {volume} {27}},\
  \bibinfo {pages} {1750178} (\bibinfo {year} {2017})},\ \Eprint
  {http://arxiv.org/abs/1011.4788} {arXiv:1011.4788 [astro-ph.CO]} \BibitemShut
  {NoStop}%
\bibitem [{\citenamefont {Braglia}\ \emph {et~al.}(2020)\citenamefont
  {Braglia}, \citenamefont {Ballardini}, \citenamefont {Emond}, \citenamefont
  {Finelli}, \citenamefont {Gumrukcuoglu}, \citenamefont {Koyama},\ and\
  \citenamefont {Paoletti}}]{Braglia:2020iik}%
  \BibitemOpen
  \bibfield  {author} {\bibinfo {author} {\bibfnamefont {M.}~\bibnamefont
  {Braglia}}, \bibinfo {author} {\bibfnamefont {M.}~\bibnamefont {Ballardini}},
  \bibinfo {author} {\bibfnamefont {W.~T.}\ \bibnamefont {Emond}}, \bibinfo
  {author} {\bibfnamefont {F.}~\bibnamefont {Finelli}}, \bibinfo {author}
  {\bibfnamefont {A.~E.}\ \bibnamefont {Gumrukcuoglu}}, \bibinfo {author}
  {\bibfnamefont {K.}~\bibnamefont {Koyama}}, \ and\ \bibinfo {author}
  {\bibfnamefont {D.}~\bibnamefont {Paoletti}},\ }\href {\doibase
  10.1103/PhysRevD.102.023529} {\bibfield  {journal} {\bibinfo  {journal}
  {Phys. Rev. D}\ }\textbf {\bibinfo {volume} {102}},\ \bibinfo {pages}
  {023529} (\bibinfo {year} {2020})},\ \Eprint
  {http://arxiv.org/abs/2004.11161} {arXiv:2004.11161 [astro-ph.CO]}
  \BibitemShut {NoStop}%
\bibitem [{\citenamefont {Raveri}(2020)}]{Raveri:2019mxg}%
  \BibitemOpen
  \bibfield  {author} {\bibinfo {author} {\bibfnamefont {M.}~\bibnamefont
  {Raveri}},\ }\href {\doibase 10.1103/PhysRevD.101.083524} {\bibfield
  {journal} {\bibinfo  {journal} {Phys. Rev. D}\ }\textbf {\bibinfo {volume}
  {101}},\ \bibinfo {pages} {083524} (\bibinfo {year} {2020})},\ \Eprint
  {http://arxiv.org/abs/1902.01366} {arXiv:1902.01366 [astro-ph.CO]}
  \BibitemShut {NoStop}%
\bibitem [{\citenamefont {Horndeski}(1974)}]{Horndeski:1974wa}%
  \BibitemOpen
  \bibfield  {author} {\bibinfo {author} {\bibfnamefont {G.~W.}\ \bibnamefont
  {Horndeski}},\ }\href {\doibase 10.1007/BF01807638} {\bibfield  {journal}
  {\bibinfo  {journal} {Int. J. Theor. Phys.}\ }\textbf {\bibinfo {volume}
  {10}},\ \bibinfo {pages} {363} (\bibinfo {year} {1974})}\BibitemShut
  {NoStop}%
\bibitem [{\citenamefont {de~Brito}\ \emph {et~al.}(2021)\citenamefont
  {de~Brito}, \citenamefont {Pereira},\ and\ \citenamefont
  {Vieira}}]{deBrito:2020xhy}%
  \BibitemOpen
  \bibfield  {author} {\bibinfo {author} {\bibfnamefont {G.~P.}\ \bibnamefont
  {de~Brito}}, \bibinfo {author} {\bibfnamefont {A.~D.}\ \bibnamefont
  {Pereira}}, \ and\ \bibinfo {author} {\bibfnamefont {A.~F.}\ \bibnamefont
  {Vieira}},\ }\href {\doibase 10.1103/PhysRevD.103.104023} {\bibfield
  {journal} {\bibinfo  {journal} {Phys. Rev. D}\ }\textbf {\bibinfo {volume}
  {103}},\ \bibinfo {pages} {104023} (\bibinfo {year} {2021})},\ \Eprint
  {http://arxiv.org/abs/2012.08904} {arXiv:2012.08904 [hep-th]} \BibitemShut
  {NoStop}%
\bibitem [{\citenamefont {Frusciante}\ and\ \citenamefont
  {Benetti}(2021)}]{Frusciante:2020gkx}%
  \BibitemOpen
  \bibfield  {author} {\bibinfo {author} {\bibfnamefont {N.}~\bibnamefont
  {Frusciante}}\ and\ \bibinfo {author} {\bibfnamefont {M.}~\bibnamefont
  {Benetti}},\ }\href {\doibase 10.1103/PhysRevD.103.104060} {\bibfield
  {journal} {\bibinfo  {journal} {Phys. Rev. D}\ }\textbf {\bibinfo {volume}
  {103}},\ \bibinfo {pages} {104060} (\bibinfo {year} {2021})},\ \Eprint
  {http://arxiv.org/abs/2005.14705} {arXiv:2005.14705 [astro-ph.CO]}
  \BibitemShut {NoStop}%
\bibitem [{\citenamefont {Benetti}\ \emph {et~al.}(2020)\citenamefont
  {Benetti}, \citenamefont {Capozziello},\ and\ \citenamefont
  {Lambiase}}]{Benetti:2020hxp}%
  \BibitemOpen
  \bibfield  {author} {\bibinfo {author} {\bibfnamefont {M.}~\bibnamefont
  {Benetti}}, \bibinfo {author} {\bibfnamefont {S.}~\bibnamefont
  {Capozziello}}, \ and\ \bibinfo {author} {\bibfnamefont {G.}~\bibnamefont
  {Lambiase}},\ }\href {\doibase 10.1093/mnras/staa3368} {\bibfield  {journal}
  {\bibinfo  {journal} {Mon. Not. Roy. Astron. Soc.}\ }\textbf {\bibinfo
  {volume} {500}},\ \bibinfo {pages} {1795} (\bibinfo {year} {2020})},\ \Eprint
  {http://arxiv.org/abs/2006.15335} {arXiv:2006.15335 [astro-ph.CO]}
  \BibitemShut {NoStop}%
\bibitem [{\citenamefont {Benetti}\ \emph
  {et~al.}(2018{\natexlab{b}})\citenamefont {Benetti}, \citenamefont {Santos~da
  Costa}, \citenamefont {Capozziello}, \citenamefont {Alcaniz},\ and\
  \citenamefont {De~Laurentis}}]{Benetti:2018zhv}%
  \BibitemOpen
  \bibfield  {author} {\bibinfo {author} {\bibfnamefont {M.}~\bibnamefont
  {Benetti}}, \bibinfo {author} {\bibfnamefont {S.}~\bibnamefont {Santos~da
  Costa}}, \bibinfo {author} {\bibfnamefont {S.}~\bibnamefont {Capozziello}},
  \bibinfo {author} {\bibfnamefont {J.~S.}\ \bibnamefont {Alcaniz}}, \ and\
  \bibinfo {author} {\bibfnamefont {M.}~\bibnamefont {De~Laurentis}},\ }\href
  {\doibase 10.1142/S0218271818500840} {\bibfield  {journal} {\bibinfo
  {journal} {Int. J. Mod. Phys. D}\ }\textbf {\bibinfo {volume} {27}},\
  \bibinfo {pages} {1850084} (\bibinfo {year} {2018}{\natexlab{b}})},\ \Eprint
  {http://arxiv.org/abs/1803.00895} {arXiv:1803.00895 [gr-qc]} \BibitemShut
  {NoStop}%
\bibitem [{\citenamefont {Rodrigues}\ \emph {et~al.}(2021)\citenamefont
  {Rodrigues}, \citenamefont {Benetti},\ and\ \citenamefont
  {Alcaniz}}]{Rodrigues:2021txa}%
  \BibitemOpen
  \bibfield  {author} {\bibinfo {author} {\bibfnamefont {J.~G.}\ \bibnamefont
  {Rodrigues}}, \bibinfo {author} {\bibfnamefont {M.}~\bibnamefont {Benetti}},
  \ and\ \bibinfo {author} {\bibfnamefont {J.~S.}\ \bibnamefont {Alcaniz}},\
  }\href {\doibase 10.1007/JHEP11(2021)091} {\bibfield  {journal} {\bibinfo
  {journal} {JHEP}\ }\textbf {\bibinfo {volume} {11}},\ \bibinfo {pages} {091}
  (\bibinfo {year} {2021})},\ \Eprint {http://arxiv.org/abs/2105.07009}
  {arXiv:2105.07009 [hep-ph]} \BibitemShut {NoStop}%
\bibitem [{\citenamefont {Chiang}\ and\ \citenamefont
  {Slosar}(2018)}]{Chiang:2018xpn}%
  \BibitemOpen
  \bibfield  {author} {\bibinfo {author} {\bibfnamefont {C.-T.}\ \bibnamefont
  {Chiang}}\ and\ \bibinfo {author} {\bibfnamefont {A.}~\bibnamefont
  {Slosar}},\ }\href@noop {} {\  (\bibinfo {year} {2018})},\ \Eprint
  {http://arxiv.org/abs/1811.03624} {arXiv:1811.03624 [astro-ph.CO]}
  \BibitemShut {NoStop}%
\bibitem [{\citenamefont {Banihashemi}\ \emph {et~al.}(2020)\citenamefont
  {Banihashemi}, \citenamefont {Khosravi},\ and\ \citenamefont
  {Shirazi}}]{Banihashemi:2018oxo}%
  \BibitemOpen
  \bibfield  {author} {\bibinfo {author} {\bibfnamefont {A.}~\bibnamefont
  {Banihashemi}}, \bibinfo {author} {\bibfnamefont {N.}~\bibnamefont
  {Khosravi}}, \ and\ \bibinfo {author} {\bibfnamefont {A.~H.}\ \bibnamefont
  {Shirazi}},\ }\href {\doibase 10.1103/PhysRevD.101.123521} {\bibfield
  {journal} {\bibinfo  {journal} {Phys. Rev. D}\ }\textbf {\bibinfo {volume}
  {101}},\ \bibinfo {pages} {123521} (\bibinfo {year} {2020})},\ \Eprint
  {http://arxiv.org/abs/1808.02472} {arXiv:1808.02472 [astro-ph.CO]}
  \BibitemShut {NoStop}%
\bibitem [{\citenamefont {Spallicci}\ \emph {et~al.}(2022)\citenamefont
  {Spallicci}, \citenamefont {Benetti},\ and\ \citenamefont
  {Capozziello}}]{Spallicci:2021kye}%
  \BibitemOpen
  \bibfield  {author} {\bibinfo {author} {\bibfnamefont {A.~D. A.~M.}\
  \bibnamefont {Spallicci}}, \bibinfo {author} {\bibfnamefont {M.}~\bibnamefont
  {Benetti}}, \ and\ \bibinfo {author} {\bibfnamefont {S.}~\bibnamefont
  {Capozziello}},\ }\href {\doibase 10.1007/s10701-021-00531-z} {\bibfield
  {journal} {\bibinfo  {journal} {Found. Phys.}\ }\textbf {\bibinfo {volume}
  {52}},\ \bibinfo {pages} {23} (\bibinfo {year} {2022})},\ \Eprint
  {http://arxiv.org/abs/2112.07359} {arXiv:2112.07359 [physics.gen-ph]}
  \BibitemShut {NoStop}%
\bibitem [{\citenamefont {Capozziello}\ \emph {et~al.}(2020)\citenamefont
  {Capozziello}, \citenamefont {Benetti},\ and\ \citenamefont
  {Spallicci}}]{Capozziello:2020nyq}%
  \BibitemOpen
  \bibfield  {author} {\bibinfo {author} {\bibfnamefont {S.}~\bibnamefont
  {Capozziello}}, \bibinfo {author} {\bibfnamefont {M.}~\bibnamefont
  {Benetti}}, \ and\ \bibinfo {author} {\bibfnamefont {A.~D. A.~M.}\
  \bibnamefont {Spallicci}},\ }\href {\doibase 10.1007/s10701-020-00356-2}
  {\bibfield  {journal} {\bibinfo  {journal} {Found. Phys.}\ }\textbf {\bibinfo
  {volume} {50}},\ \bibinfo {pages} {893} (\bibinfo {year} {2020})},\ \Eprint
  {http://arxiv.org/abs/2007.00462} {arXiv:2007.00462 [gr-qc]} \BibitemShut
  {NoStop}%
\bibitem [{\citenamefont {Benetti}\ and\ \citenamefont
  {Capozziello}(2019)}]{Benetti:2019gmo}%
  \BibitemOpen
  \bibfield  {author} {\bibinfo {author} {\bibfnamefont {M.}~\bibnamefont
  {Benetti}}\ and\ \bibinfo {author} {\bibfnamefont {S.}~\bibnamefont
  {Capozziello}},\ }\href {\doibase 10.1088/1475-7516/2019/12/008} {\bibfield
  {journal} {\bibinfo  {journal} {JCAP}\ }\textbf {\bibinfo {volume} {12}},\
  \bibinfo {pages} {008} (\bibinfo {year} {2019})},\ \Eprint
  {http://arxiv.org/abs/1910.09975} {arXiv:1910.09975 [astro-ph.CO]}
  \BibitemShut {NoStop}%
\bibitem [{\citenamefont {Freedman}\ \emph {et~al.}(2020)\citenamefont
  {Freedman}, \citenamefont {Madore}, \citenamefont {Hoyt}, \citenamefont
  {Jang}, \citenamefont {Beaton}, \citenamefont {Lee}, \citenamefont {Monson},
  \citenamefont {Neeley},\ and\ \citenamefont {Rich}}]{Freedman:2020dne}%
  \BibitemOpen
  \bibfield  {author} {\bibinfo {author} {\bibfnamefont {W.~L.}\ \bibnamefont
  {Freedman}}, \bibinfo {author} {\bibfnamefont {B.~F.}\ \bibnamefont
  {Madore}}, \bibinfo {author} {\bibfnamefont {T.}~\bibnamefont {Hoyt}},
  \bibinfo {author} {\bibfnamefont {I.~S.}\ \bibnamefont {Jang}}, \bibinfo
  {author} {\bibfnamefont {R.}~\bibnamefont {Beaton}}, \bibinfo {author}
  {\bibfnamefont {M.~G.}\ \bibnamefont {Lee}}, \bibinfo {author} {\bibfnamefont
  {A.}~\bibnamefont {Monson}}, \bibinfo {author} {\bibfnamefont
  {J.}~\bibnamefont {Neeley}}, \ and\ \bibinfo {author} {\bibfnamefont
  {J.}~\bibnamefont {Rich}},\ }\href {\doibase 10.3847/1538-4357/ab7339} {\
  (\bibinfo {year} {2020}),\ 10.3847/1538-4357/ab7339},\ \Eprint
  {http://arxiv.org/abs/2002.01550} {arXiv:2002.01550 [astro-ph.GA]}
  \BibitemShut {NoStop}%
\bibitem [{\citenamefont {Benevento}\ \emph {et~al.}(2020)\citenamefont
  {Benevento}, \citenamefont {Hu},\ and\ \citenamefont
  {Raveri}}]{Benevento:2020fev}%
  \BibitemOpen
  \bibfield  {author} {\bibinfo {author} {\bibfnamefont {G.}~\bibnamefont
  {Benevento}}, \bibinfo {author} {\bibfnamefont {W.}~\bibnamefont {Hu}}, \
  and\ \bibinfo {author} {\bibfnamefont {M.}~\bibnamefont {Raveri}},\ }\href
  {\doibase 10.1103/PhysRevD.101.103517} {\bibfield  {journal} {\bibinfo
  {journal} {Phys. Rev. D}\ }\textbf {\bibinfo {volume} {101}},\ \bibinfo
  {pages} {103517} (\bibinfo {year} {2020})},\ \Eprint
  {http://arxiv.org/abs/2002.11707} {arXiv:2002.11707 [astro-ph.CO]}
  \BibitemShut {NoStop}%
\bibitem [{\citenamefont {Banerjee}\ \emph {et~al.}(2021)\citenamefont
  {Banerjee}, \citenamefont {Cai}, \citenamefont {Heisenberg}, \citenamefont
  {Colg\'ain}, \citenamefont {Sheikh-Jabbari},\ and\ \citenamefont
  {Yang}}]{Banerjee:2020xcn}%
  \BibitemOpen
  \bibfield  {author} {\bibinfo {author} {\bibfnamefont {A.}~\bibnamefont
  {Banerjee}}, \bibinfo {author} {\bibfnamefont {H.}~\bibnamefont {Cai}},
  \bibinfo {author} {\bibfnamefont {L.}~\bibnamefont {Heisenberg}}, \bibinfo
  {author} {\bibfnamefont {E.~O.}\ \bibnamefont {Colg\'ain}}, \bibinfo {author}
  {\bibfnamefont {M.~M.}\ \bibnamefont {Sheikh-Jabbari}}, \ and\ \bibinfo
  {author} {\bibfnamefont {T.}~\bibnamefont {Yang}},\ }\href {\doibase
  10.1103/PhysRevD.103.L081305} {\bibfield  {journal} {\bibinfo  {journal}
  {Phys. Rev. D}\ }\textbf {\bibinfo {volume} {103}},\ \bibinfo {pages}
  {L081305} (\bibinfo {year} {2021})},\ \Eprint
  {http://arxiv.org/abs/2006.00244} {arXiv:2006.00244 [astro-ph.CO]}
  \BibitemShut {NoStop}%
\bibitem [{\citenamefont {Lee}\ \emph {et~al.}(2022)\citenamefont {Lee},
  \citenamefont {Lee}, \citenamefont {Colg\'ain}, \citenamefont
  {Sheikh-Jabbari},\ and\ \citenamefont {Thakur}}]{Lee:2022cyh}%
  \BibitemOpen
  \bibfield  {author} {\bibinfo {author} {\bibfnamefont {B.-H.}\ \bibnamefont
  {Lee}}, \bibinfo {author} {\bibfnamefont {W.}~\bibnamefont {Lee}}, \bibinfo
  {author} {\bibfnamefont {E.~O.}\ \bibnamefont {Colg\'ain}}, \bibinfo {author}
  {\bibfnamefont {M.~M.}\ \bibnamefont {Sheikh-Jabbari}}, \ and\ \bibinfo
  {author} {\bibfnamefont {S.}~\bibnamefont {Thakur}},\ }\href {\doibase
  10.1088/1475-7516/2022/04/004} {\bibfield  {journal} {\bibinfo  {journal}
  {JCAP}\ }\textbf {\bibinfo {volume} {04}},\ \bibinfo {pages} {004} (\bibinfo
  {year} {2022})},\ \Eprint {http://arxiv.org/abs/2202.03906} {arXiv:2202.03906
  [astro-ph.CO]} \BibitemShut {NoStop}%
\bibitem [{\citenamefont {Vagnozzi}(2020)}]{Vagnozzi:2019ezj}%
  \BibitemOpen
  \bibfield  {author} {\bibinfo {author} {\bibfnamefont {S.}~\bibnamefont
  {Vagnozzi}},\ }\href {\doibase 10.1103/PhysRevD.102.023518} {\bibfield
  {journal} {\bibinfo  {journal} {Phys. Rev. D}\ }\textbf {\bibinfo {volume}
  {102}},\ \bibinfo {pages} {023518} (\bibinfo {year} {2020})},\ \Eprint
  {http://arxiv.org/abs/1907.07569} {arXiv:1907.07569 [astro-ph.CO]}
  \BibitemShut {NoStop}%
\bibitem [{\citenamefont {Krishnan}\ \emph {et~al.}(2021)\citenamefont
  {Krishnan}, \citenamefont {Colg\'ain}, \citenamefont {Sheikh-Jabbari},\ and\
  \citenamefont {Yang}}]{Krishnan:2020vaf}%
  \BibitemOpen
  \bibfield  {author} {\bibinfo {author} {\bibfnamefont {C.}~\bibnamefont
  {Krishnan}}, \bibinfo {author} {\bibfnamefont {E.~O.}\ \bibnamefont
  {Colg\'ain}}, \bibinfo {author} {\bibfnamefont {M.~M.}\ \bibnamefont
  {Sheikh-Jabbari}}, \ and\ \bibinfo {author} {\bibfnamefont {T.}~\bibnamefont
  {Yang}},\ }\href {\doibase 10.1103/PhysRevD.103.103509} {\bibfield  {journal}
  {\bibinfo  {journal} {Phys. Rev. D}\ }\textbf {\bibinfo {volume} {103}},\
  \bibinfo {pages} {103509} (\bibinfo {year} {2021})},\ \Eprint
  {http://arxiv.org/abs/2011.02858} {arXiv:2011.02858 [astro-ph.CO]}
  \BibitemShut {NoStop}%
\bibitem [{\citenamefont {Colg\'ain}\ \emph
  {et~al.}(2022{\natexlab{a}})\citenamefont {Colg\'ain}, \citenamefont
  {Sheikh-Jabbari}, \citenamefont {Solomon}, \citenamefont {Dainotti},\ and\
  \citenamefont {Stojkovic}}]{Colgain:2022rxy}%
  \BibitemOpen
  \bibfield  {author} {\bibinfo {author} {\bibfnamefont {E.~O.}\ \bibnamefont
  {Colg\'ain}}, \bibinfo {author} {\bibfnamefont {M.~M.}\ \bibnamefont
  {Sheikh-Jabbari}}, \bibinfo {author} {\bibfnamefont {R.}~\bibnamefont
  {Solomon}}, \bibinfo {author} {\bibfnamefont {M.~G.}\ \bibnamefont
  {Dainotti}}, \ and\ \bibinfo {author} {\bibfnamefont {D.}~\bibnamefont
  {Stojkovic}},\ }\href@noop {} {\  (\bibinfo {year} {2022}{\natexlab{a}})},\
  \Eprint {http://arxiv.org/abs/2206.11447} {arXiv:2206.11447 [astro-ph.CO]}
  \BibitemShut {NoStop}%
\bibitem [{\citenamefont {Colg\'ain}\ \emph
  {et~al.}(2022{\natexlab{b}})\citenamefont {Colg\'ain}, \citenamefont
  {Sheikh-Jabbari}, \citenamefont {Solomon}, \citenamefont {Bargiacchi},
  \citenamefont {Capozziello}, \citenamefont {Dainotti},\ and\ \citenamefont
  {Stojkovic}}]{Colgain:2022nlb}%
  \BibitemOpen
  \bibfield  {author} {\bibinfo {author} {\bibfnamefont {E.~O.}\ \bibnamefont
  {Colg\'ain}}, \bibinfo {author} {\bibfnamefont {M.~M.}\ \bibnamefont
  {Sheikh-Jabbari}}, \bibinfo {author} {\bibfnamefont {R.}~\bibnamefont
  {Solomon}}, \bibinfo {author} {\bibfnamefont {G.}~\bibnamefont {Bargiacchi}},
  \bibinfo {author} {\bibfnamefont {S.}~\bibnamefont {Capozziello}}, \bibinfo
  {author} {\bibfnamefont {M.~G.}\ \bibnamefont {Dainotti}}, \ and\ \bibinfo
  {author} {\bibfnamefont {D.}~\bibnamefont {Stojkovic}},\ }\href {\doibase
  10.1103/PhysRevD.106.L041301} {\bibfield  {journal} {\bibinfo  {journal}
  {Phys. Rev. D}\ }\textbf {\bibinfo {volume} {106}},\ \bibinfo {pages}
  {L041301} (\bibinfo {year} {2022}{\natexlab{b}})},\ \Eprint
  {http://arxiv.org/abs/2203.10558} {arXiv:2203.10558 [astro-ph.CO]}
  \BibitemShut {NoStop}%
\bibitem [{\citenamefont {Jedamzik}\ \emph {et~al.}(2021)\citenamefont
  {Jedamzik}, \citenamefont {Pogosian},\ and\ \citenamefont
  {Zhao}}]{Jedamzik:2020zmd}%
  \BibitemOpen
  \bibfield  {author} {\bibinfo {author} {\bibfnamefont {K.}~\bibnamefont
  {Jedamzik}}, \bibinfo {author} {\bibfnamefont {L.}~\bibnamefont {Pogosian}},
  \ and\ \bibinfo {author} {\bibfnamefont {G.-B.}\ \bibnamefont {Zhao}},\
  }\href {\doibase 10.1038/s42005-021-00628-x} {\bibfield  {journal} {\bibinfo
  {journal} {Commun. in Phys.}\ }\textbf {\bibinfo {volume} {4}},\ \bibinfo
  {pages} {123} (\bibinfo {year} {2021})},\ \Eprint
  {http://arxiv.org/abs/2010.04158} {arXiv:2010.04158 [astro-ph.CO]}
  \BibitemShut {NoStop}%
\bibitem [{\citenamefont {Naidoo}\ \emph {et~al.}(2022)\citenamefont {Naidoo},
  \citenamefont {Jaber}, \citenamefont {Hellwing},\ and\ \citenamefont
  {Bilicki}}]{Naidoo:2022rda}%
  \BibitemOpen
  \bibfield  {author} {\bibinfo {author} {\bibfnamefont {K.}~\bibnamefont
  {Naidoo}}, \bibinfo {author} {\bibfnamefont {M.}~\bibnamefont {Jaber}},
  \bibinfo {author} {\bibfnamefont {W.~A.}\ \bibnamefont {Hellwing}}, \ and\
  \bibinfo {author} {\bibfnamefont {M.}~\bibnamefont {Bilicki}},\ }\href@noop
  {} {\  (\bibinfo {year} {2022})},\ \Eprint {http://arxiv.org/abs/2209.08102}
  {arXiv:2209.08102 [astro-ph.CO]} \BibitemShut {NoStop}%
\bibitem [{\citenamefont {Heymans}\ \emph {et~al.}(2021)\citenamefont {Heymans}
  \emph {et~al.}}]{Heymans:2020gsg}%
  \BibitemOpen
  \bibfield  {author} {\bibinfo {author} {\bibfnamefont {C.}~\bibnamefont
  {Heymans}} \emph {et~al.},\ }\href {\doibase 10.1051/0004-6361/202039063}
  {\bibfield  {journal} {\bibinfo  {journal} {Astron. Astrophys.}\ }\textbf
  {\bibinfo {volume} {646}},\ \bibinfo {pages} {A140} (\bibinfo {year}
  {2021})},\ \Eprint {http://arxiv.org/abs/2007.15632} {arXiv:2007.15632
  [astro-ph.CO]} \BibitemShut {NoStop}%
\bibitem [{\citenamefont {Amon}\ \emph {et~al.}(2022)\citenamefont {Amon} \emph
  {et~al.}}]{DES:2021bvc}%
  \BibitemOpen
  \bibfield  {author} {\bibinfo {author} {\bibfnamefont {A.}~\bibnamefont
  {Amon}} \emph {et~al.} (\bibinfo {collaboration} {DES}),\ }\href {\doibase
  10.1103/PhysRevD.105.023514} {\bibfield  {journal} {\bibinfo  {journal}
  {Phys. Rev. D}\ }\textbf {\bibinfo {volume} {105}},\ \bibinfo {pages}
  {023514} (\bibinfo {year} {2022})},\ \Eprint
  {http://arxiv.org/abs/2105.13543} {arXiv:2105.13543 [astro-ph.CO]}
  \BibitemShut {NoStop}%
\bibitem [{\citenamefont {Nunes}\ and\ \citenamefont
  {Vagnozzi}(2021)}]{Nunes:2021ipq}%
  \BibitemOpen
  \bibfield  {author} {\bibinfo {author} {\bibfnamefont {R.~C.}\ \bibnamefont
  {Nunes}}\ and\ \bibinfo {author} {\bibfnamefont {S.}~\bibnamefont
  {Vagnozzi}},\ }\href {\doibase 10.1093/mnras/stab1613} {\bibfield  {journal}
  {\bibinfo  {journal} {Mon. Not. Roy. Astron. Soc.}\ }\textbf {\bibinfo
  {volume} {505}},\ \bibinfo {pages} {5427} (\bibinfo {year} {2021})},\ \Eprint
  {http://arxiv.org/abs/2106.01208} {arXiv:2106.01208 [astro-ph.CO]}
  \BibitemShut {NoStop}%
\bibitem [{\citenamefont {Zhang}\ and\ \citenamefont
  {Huang}(2021)}]{Zhang:2020uan}%
  \BibitemOpen
  \bibfield  {author} {\bibinfo {author} {\bibfnamefont {X.}~\bibnamefont
  {Zhang}}\ and\ \bibinfo {author} {\bibfnamefont {Q.-G.}\ \bibnamefont
  {Huang}},\ }\href {\doibase 10.1103/PhysRevD.103.043513} {\bibfield
  {journal} {\bibinfo  {journal} {Phys. Rev. D}\ }\textbf {\bibinfo {volume}
  {103}},\ \bibinfo {pages} {043513} (\bibinfo {year} {2021})},\ \Eprint
  {http://arxiv.org/abs/2006.16692} {arXiv:2006.16692 [astro-ph.CO]}
  \BibitemShut {NoStop}%
\bibitem [{\citenamefont {Bernal}\ \emph {et~al.}(2016)\citenamefont {Bernal},
  \citenamefont {Verde},\ and\ \citenamefont {Riess}}]{Bernal:2016gxb}%
  \BibitemOpen
  \bibfield  {author} {\bibinfo {author} {\bibfnamefont {J.~L.}\ \bibnamefont
  {Bernal}}, \bibinfo {author} {\bibfnamefont {L.}~\bibnamefont {Verde}}, \
  and\ \bibinfo {author} {\bibfnamefont {A.~G.}\ \bibnamefont {Riess}},\ }\href
  {\doibase 10.1088/1475-7516/2016/10/019} {\bibfield  {journal} {\bibinfo
  {journal} {JCAP}\ }\textbf {\bibinfo {volume} {10}},\ \bibinfo {pages} {019}
  (\bibinfo {year} {2016})},\ \Eprint {http://arxiv.org/abs/1607.05617}
  {arXiv:1607.05617 [astro-ph.CO]} \BibitemShut {NoStop}%
\bibitem [{\citenamefont {Arendse}\ \emph {et~al.}(2020)\citenamefont {Arendse}
  \emph {et~al.}}]{Arendse:2019hev}%
  \BibitemOpen
  \bibfield  {author} {\bibinfo {author} {\bibfnamefont {N.}~\bibnamefont
  {Arendse}} \emph {et~al.},\ }\href {\doibase 10.1051/0004-6361/201936720}
  {\bibfield  {journal} {\bibinfo  {journal} {Astron. Astrophys.}\ }\textbf
  {\bibinfo {volume} {639}},\ \bibinfo {pages} {A57} (\bibinfo {year}
  {2020})},\ \Eprint {http://arxiv.org/abs/1909.07986} {arXiv:1909.07986
  [astro-ph.CO]} \BibitemShut {NoStop}%
\bibitem [{\citenamefont {Bargiacchi}\ \emph {et~al.}(2021)\citenamefont
  {Bargiacchi}, \citenamefont {Benetti}, \citenamefont {Capozziello},
  \citenamefont {Lusso}, \citenamefont {Risaliti},\ and\ \citenamefont
  {Signorini}}]{Bargiacchi:2021hdp}%
  \BibitemOpen
  \bibfield  {author} {\bibinfo {author} {\bibfnamefont {G.}~\bibnamefont
  {Bargiacchi}}, \bibinfo {author} {\bibfnamefont {M.}~\bibnamefont {Benetti}},
  \bibinfo {author} {\bibfnamefont {S.}~\bibnamefont {Capozziello}}, \bibinfo
  {author} {\bibfnamefont {E.}~\bibnamefont {Lusso}}, \bibinfo {author}
  {\bibfnamefont {G.}~\bibnamefont {Risaliti}}, \ and\ \bibinfo {author}
  {\bibfnamefont {M.}~\bibnamefont {Signorini}},\ }\href@noop {} {\  (\bibinfo
  {year} {2021})},\ \Eprint {http://arxiv.org/abs/2111.02420} {arXiv:2111.02420
  [astro-ph.CO]} \BibitemShut {NoStop}%
\bibitem [{\citenamefont {Joudaki}\ \emph {et~al.}(2017)\citenamefont {Joudaki}
  \emph {et~al.}}]{Joudaki:2016kym}%
  \BibitemOpen
  \bibfield  {author} {\bibinfo {author} {\bibfnamefont {S.}~\bibnamefont
  {Joudaki}} \emph {et~al.},\ }\href {\doibase 10.1093/mnras/stx998} {\bibfield
   {journal} {\bibinfo  {journal} {Mon. Not. Roy. Astron. Soc.}\ }\textbf
  {\bibinfo {volume} {471}},\ \bibinfo {pages} {1259} (\bibinfo {year}
  {2017})},\ \Eprint {http://arxiv.org/abs/1610.04606} {arXiv:1610.04606
  [astro-ph.CO]} \BibitemShut {NoStop}%
\bibitem [{\citenamefont {Zhang}\ and\ \citenamefont
  {Huang}(2019)}]{Zhang:2018air}%
  \BibitemOpen
  \bibfield  {author} {\bibinfo {author} {\bibfnamefont {X.}~\bibnamefont
  {Zhang}}\ and\ \bibinfo {author} {\bibfnamefont {Q.-G.}\ \bibnamefont
  {Huang}},\ }\href {\doibase 10.1088/0253-6102/71/7/826} {\bibfield  {journal}
  {\bibinfo  {journal} {Commun. Theor. Phys.}\ }\textbf {\bibinfo {volume}
  {71}},\ \bibinfo {pages} {826} (\bibinfo {year} {2019})},\ \Eprint
  {http://arxiv.org/abs/1812.01877} {arXiv:1812.01877 [astro-ph.CO]}
  \BibitemShut {NoStop}%
\bibitem [{\citenamefont {Visinelli}\ \emph {et~al.}(2019)\citenamefont
  {Visinelli}, \citenamefont {Vagnozzi},\ and\ \citenamefont
  {Danielsson}}]{Visinelli:2019qqu}%
  \BibitemOpen
  \bibfield  {author} {\bibinfo {author} {\bibfnamefont {L.}~\bibnamefont
  {Visinelli}}, \bibinfo {author} {\bibfnamefont {S.}~\bibnamefont {Vagnozzi}},
  \ and\ \bibinfo {author} {\bibfnamefont {U.}~\bibnamefont {Danielsson}},\
  }\href {\doibase 10.3390/sym11081035} {\bibfield  {journal} {\bibinfo
  {journal} {Symmetry}\ }\textbf {\bibinfo {volume} {11}},\ \bibinfo {pages}
  {1035} (\bibinfo {year} {2019})},\ \Eprint {http://arxiv.org/abs/1907.07953}
  {arXiv:1907.07953 [astro-ph.CO]} \BibitemShut {NoStop}%
\bibitem [{\citenamefont {Di~Valentino}\ \emph
  {et~al.}(2020{\natexlab{b}})\citenamefont {Di~Valentino}, \citenamefont
  {Melchiorri}, \citenamefont {Mena},\ and\ \citenamefont
  {Vagnozzi}}]{DiValentino:2019jae}%
  \BibitemOpen
  \bibfield  {author} {\bibinfo {author} {\bibfnamefont {E.}~\bibnamefont
  {Di~Valentino}}, \bibinfo {author} {\bibfnamefont {A.}~\bibnamefont
  {Melchiorri}}, \bibinfo {author} {\bibfnamefont {O.}~\bibnamefont {Mena}}, \
  and\ \bibinfo {author} {\bibfnamefont {S.}~\bibnamefont {Vagnozzi}},\ }\href
  {\doibase 10.1103/PhysRevD.101.063502} {\bibfield  {journal} {\bibinfo
  {journal} {Phys. Rev. D}\ }\textbf {\bibinfo {volume} {101}},\ \bibinfo
  {pages} {063502} (\bibinfo {year} {2020}{\natexlab{b}})},\ \Eprint
  {http://arxiv.org/abs/1910.09853} {arXiv:1910.09853 [astro-ph.CO]}
  \BibitemShut {NoStop}%
\bibitem [{\citenamefont {Keeley}\ \emph {et~al.}(2019)\citenamefont {Keeley},
  \citenamefont {Joudaki}, \citenamefont {Kaplinghat},\ and\ \citenamefont
  {Kirkby}}]{Keeley:2019esp}%
  \BibitemOpen
  \bibfield  {author} {\bibinfo {author} {\bibfnamefont {R.~E.}\ \bibnamefont
  {Keeley}}, \bibinfo {author} {\bibfnamefont {S.}~\bibnamefont {Joudaki}},
  \bibinfo {author} {\bibfnamefont {M.}~\bibnamefont {Kaplinghat}}, \ and\
  \bibinfo {author} {\bibfnamefont {D.}~\bibnamefont {Kirkby}},\ }\href
  {\doibase 10.1088/1475-7516/2019/12/035} {\bibfield  {journal} {\bibinfo
  {journal} {JCAP}\ }\textbf {\bibinfo {volume} {12}},\ \bibinfo {pages} {035}
  (\bibinfo {year} {2019})},\ \Eprint {http://arxiv.org/abs/1905.10198}
  {arXiv:1905.10198 [astro-ph.CO]} \BibitemShut {NoStop}%
\bibitem [{\citenamefont {Abdalla}\ \emph {et~al.}(2022)\citenamefont {Abdalla}
  \emph {et~al.}}]{Abdalla:2022yfr}%
  \BibitemOpen
  \bibfield  {author} {\bibinfo {author} {\bibfnamefont {E.}~\bibnamefont
  {Abdalla}} \emph {et~al.},\ }\href {\doibase 10.1016/j.jheap.2022.04.002}
  {\bibfield  {journal} {\bibinfo  {journal} {JHEAp}\ }\textbf {\bibinfo
  {volume} {34}},\ \bibinfo {pages} {49} (\bibinfo {year} {2022})},\ \Eprint
  {http://arxiv.org/abs/2203.06142} {arXiv:2203.06142 [astro-ph.CO]}
  \BibitemShut {NoStop}%
\bibitem [{\citenamefont {Benetti}\ \emph {et~al.}(2022)\citenamefont
  {Benetti}, \citenamefont {Graef},\ and\ \citenamefont
  {Vagnozzi}}]{Benetti:2021uea}%
  \BibitemOpen
  \bibfield  {author} {\bibinfo {author} {\bibfnamefont {M.}~\bibnamefont
  {Benetti}}, \bibinfo {author} {\bibfnamefont {L.~L.}\ \bibnamefont {Graef}},
  \ and\ \bibinfo {author} {\bibfnamefont {S.}~\bibnamefont {Vagnozzi}},\
  }\href {\doibase 10.1103/PhysRevD.105.043520} {\bibfield  {journal} {\bibinfo
   {journal} {Phys. Rev. D}\ }\textbf {\bibinfo {volume} {105}},\ \bibinfo
  {pages} {043520} (\bibinfo {year} {2022})},\ \Eprint
  {http://arxiv.org/abs/2111.04758} {arXiv:2111.04758 [astro-ph.CO]}
  \BibitemShut {NoStop}%
\bibitem [{\citenamefont {Maggiore}(2000)}]{Maggiore:1999vm}%
  \BibitemOpen
  \bibfield  {author} {\bibinfo {author} {\bibfnamefont {M.}~\bibnamefont
  {Maggiore}},\ }\href {\doibase 10.1016/S0370-1573(99)00102-7} {\bibfield
  {journal} {\bibinfo  {journal} {Phys. Rept.}\ }\textbf {\bibinfo {volume}
  {331}},\ \bibinfo {pages} {283} (\bibinfo {year} {2000})},\ \Eprint
  {http://arxiv.org/abs/gr-qc/9909001} {arXiv:gr-qc/9909001} \BibitemShut
  {NoStop}%
\bibitem [{\citenamefont {Knox}\ and\ \citenamefont
  {Millea}(2020)}]{Knox:2019rjx}%
  \BibitemOpen
  \bibfield  {author} {\bibinfo {author} {\bibfnamefont {L.}~\bibnamefont
  {Knox}}\ and\ \bibinfo {author} {\bibfnamefont {M.}~\bibnamefont {Millea}},\
  }\href {\doibase 10.1103/PhysRevD.101.043533} {\bibfield  {journal} {\bibinfo
   {journal} {Phys. Rev. D}\ }\textbf {\bibinfo {volume} {101}},\ \bibinfo
  {pages} {043533} (\bibinfo {year} {2020})},\ \Eprint
  {http://arxiv.org/abs/1908.03663} {arXiv:1908.03663 [astro-ph.CO]}
  \BibitemShut {NoStop}%
\bibitem [{\citenamefont {Lewis}\ and\ \citenamefont
  {Bridle}(2002)}]{Lewis:2002ah}%
  \BibitemOpen
  \bibfield  {author} {\bibinfo {author} {\bibfnamefont {A.}~\bibnamefont
  {Lewis}}\ and\ \bibinfo {author} {\bibfnamefont {S.}~\bibnamefont {Bridle}},\
  }\href {\doibase 10.1103/PhysRevD.66.103511} {\bibfield  {journal} {\bibinfo
  {journal} {Phys. Rev. D}\ }\textbf {\bibinfo {volume} {66}},\ \bibinfo
  {pages} {103511} (\bibinfo {year} {2002})},\ \Eprint
  {http://arxiv.org/abs/astro-ph/0205436} {arXiv:astro-ph/0205436} \BibitemShut
  {NoStop}%
\bibitem [{\citenamefont {Aghanim}\ \emph
  {et~al.}(2020{\natexlab{b}})\citenamefont {Aghanim} \emph
  {et~al.}}]{Planck:2019nip}%
  \BibitemOpen
  \bibfield  {author} {\bibinfo {author} {\bibfnamefont {N.}~\bibnamefont
  {Aghanim}} \emph {et~al.} (\bibinfo {collaboration} {Planck}),\ }\href
  {\doibase 10.1051/0004-6361/201936386} {\bibfield  {journal} {\bibinfo
  {journal} {Astron. Astrophys.}\ }\textbf {\bibinfo {volume} {641}},\ \bibinfo
  {pages} {A5} (\bibinfo {year} {2020}{\natexlab{b}})},\ \Eprint
  {http://arxiv.org/abs/1907.12875} {arXiv:1907.12875 [astro-ph.CO]}
  \BibitemShut {NoStop}%
\bibitem [{\citenamefont {Aghanim}\ \emph
  {et~al.}(2020{\natexlab{c}})\citenamefont {Aghanim} \emph
  {et~al.}}]{Planck:2018lbu}%
  \BibitemOpen
  \bibfield  {author} {\bibinfo {author} {\bibfnamefont {N.}~\bibnamefont
  {Aghanim}} \emph {et~al.} (\bibinfo {collaboration} {Planck}),\ }\href
  {\doibase 10.1051/0004-6361/201833886} {\bibfield  {journal} {\bibinfo
  {journal} {Astron. Astrophys.}\ }\textbf {\bibinfo {volume} {641}},\ \bibinfo
  {pages} {A8} (\bibinfo {year} {2020}{\natexlab{c}})},\ \Eprint
  {http://arxiv.org/abs/1807.06210} {arXiv:1807.06210 [astro-ph.CO]}
  \BibitemShut {NoStop}%
\bibitem [{\citenamefont {Beutler}\ \emph {et~al.}(2011)\citenamefont
  {Beutler}, \citenamefont {Blake}, \citenamefont {Colless}, \citenamefont
  {Jones}, \citenamefont {Staveley-Smith}, \citenamefont {Campbell},
  \citenamefont {Parker}, \citenamefont {Saunders},\ and\ \citenamefont
  {Watson}}]{Beutler:2011hx}%
  \BibitemOpen
  \bibfield  {author} {\bibinfo {author} {\bibfnamefont {F.}~\bibnamefont
  {Beutler}}, \bibinfo {author} {\bibfnamefont {C.}~\bibnamefont {Blake}},
  \bibinfo {author} {\bibfnamefont {M.}~\bibnamefont {Colless}}, \bibinfo
  {author} {\bibfnamefont {D.~H.}\ \bibnamefont {Jones}}, \bibinfo {author}
  {\bibfnamefont {L.}~\bibnamefont {Staveley-Smith}}, \bibinfo {author}
  {\bibfnamefont {L.}~\bibnamefont {Campbell}}, \bibinfo {author}
  {\bibfnamefont {Q.}~\bibnamefont {Parker}}, \bibinfo {author} {\bibfnamefont
  {W.}~\bibnamefont {Saunders}}, \ and\ \bibinfo {author} {\bibfnamefont
  {F.}~\bibnamefont {Watson}},\ }\href {\doibase
  10.1111/j.1365-2966.2011.19250.x} {\bibfield  {journal} {\bibinfo  {journal}
  {Mon. Not. Roy. Astron. Soc.}\ }\textbf {\bibinfo {volume} {416}},\ \bibinfo
  {pages} {3017} (\bibinfo {year} {2011})},\ \Eprint
  {http://arxiv.org/abs/1106.3366} {arXiv:1106.3366 [astro-ph.CO]} \BibitemShut
  {NoStop}%
\bibitem [{\citenamefont {Ross}\ \emph {et~al.}(2015)\citenamefont {Ross},
  \citenamefont {Samushia}, \citenamefont {Howlett}, \citenamefont {Percival},
  \citenamefont {Burden},\ and\ \citenamefont {Manera}}]{Ross:2014qpa}%
  \BibitemOpen
  \bibfield  {author} {\bibinfo {author} {\bibfnamefont {A.~J.}\ \bibnamefont
  {Ross}}, \bibinfo {author} {\bibfnamefont {L.}~\bibnamefont {Samushia}},
  \bibinfo {author} {\bibfnamefont {C.}~\bibnamefont {Howlett}}, \bibinfo
  {author} {\bibfnamefont {W.~J.}\ \bibnamefont {Percival}}, \bibinfo {author}
  {\bibfnamefont {A.}~\bibnamefont {Burden}}, \ and\ \bibinfo {author}
  {\bibfnamefont {M.}~\bibnamefont {Manera}},\ }\href {\doibase
  10.1093/mnras/stv154} {\bibfield  {journal} {\bibinfo  {journal} {Mon. Not.
  Roy. Astron. Soc.}\ }\textbf {\bibinfo {volume} {449}},\ \bibinfo {pages}
  {835} (\bibinfo {year} {2015})},\ \Eprint {http://arxiv.org/abs/1409.3242}
  {arXiv:1409.3242 [astro-ph.CO]} \BibitemShut {NoStop}%
\bibitem [{\citenamefont {Alam}\ \emph {et~al.}(2017)\citenamefont {Alam} \emph
  {et~al.}}]{BOSS:2016wmc}%
  \BibitemOpen
  \bibfield  {author} {\bibinfo {author} {\bibfnamefont {S.}~\bibnamefont
  {Alam}} \emph {et~al.} (\bibinfo {collaboration} {BOSS}),\ }\href {\doibase
  10.1093/mnras/stx721} {\bibfield  {journal} {\bibinfo  {journal} {Mon. Not.
  Roy. Astron. Soc.}\ }\textbf {\bibinfo {volume} {470}},\ \bibinfo {pages}
  {2617} (\bibinfo {year} {2017})},\ \Eprint {http://arxiv.org/abs/1607.03155}
  {arXiv:1607.03155 [astro-ph.CO]} \BibitemShut {NoStop}%
\bibitem [{\citenamefont {Scolnic}\ \emph {et~al.}(2018)\citenamefont {Scolnic}
  \emph {et~al.}}]{Scolnic:2017caz}%
  \BibitemOpen
  \bibfield  {author} {\bibinfo {author} {\bibfnamefont {D.~M.}\ \bibnamefont
  {Scolnic}} \emph {et~al.},\ }\href {\doibase 10.3847/1538-4357/aab9bb}
  {\bibfield  {journal} {\bibinfo  {journal} {Astrophys. J.}\ }\textbf
  {\bibinfo {volume} {859}},\ \bibinfo {pages} {101} (\bibinfo {year}
  {2018})},\ \Eprint {http://arxiv.org/abs/1710.00845} {arXiv:1710.00845
  [astro-ph.CO]} \BibitemShut {NoStop}%
\bibitem [{\citenamefont {Riess}\ \emph {et~al.}(2018)\citenamefont {Riess}
  \emph {et~al.}}]{Riess:2018uxu}%
  \BibitemOpen
  \bibfield  {author} {\bibinfo {author} {\bibfnamefont {A.~G.}\ \bibnamefont
  {Riess}} \emph {et~al.},\ }\href {\doibase 10.3847/1538-4357/aaadb7}
  {\bibfield  {journal} {\bibinfo  {journal} {Astrophys. J.}\ }\textbf
  {\bibinfo {volume} {855}},\ \bibinfo {pages} {136} (\bibinfo {year}
  {2018})},\ \Eprint {http://arxiv.org/abs/1801.01120} {arXiv:1801.01120
  [astro-ph.SR]} \BibitemShut {NoStop}%
\bibitem [{\citenamefont {Aver}\ \emph {et~al.}(2015)\citenamefont {Aver},
  \citenamefont {Olive},\ and\ \citenamefont {Skillman}}]{Aver:2015iza}%
  \BibitemOpen
  \bibfield  {author} {\bibinfo {author} {\bibfnamefont {E.}~\bibnamefont
  {Aver}}, \bibinfo {author} {\bibfnamefont {K.~A.}\ \bibnamefont {Olive}}, \
  and\ \bibinfo {author} {\bibfnamefont {E.~D.}\ \bibnamefont {Skillman}},\
  }\href {\doibase 10.1088/1475-7516/2015/07/011} {\bibfield  {journal}
  {\bibinfo  {journal} {JCAP}\ }\textbf {\bibinfo {volume} {07}},\ \bibinfo
  {pages} {011} (\bibinfo {year} {2015})},\ \Eprint
  {http://arxiv.org/abs/1503.08146} {arXiv:1503.08146 [astro-ph.CO]}
  \BibitemShut {NoStop}%
\bibitem [{\citenamefont {Asgari}\ \emph {et~al.}(2021)\citenamefont {Asgari}
  \emph {et~al.}}]{KiDS:2020suj}%
  \BibitemOpen
  \bibfield  {author} {\bibinfo {author} {\bibfnamefont {M.}~\bibnamefont
  {Asgari}} \emph {et~al.} (\bibinfo {collaboration} {KiDS}),\ }\href {\doibase
  10.1051/0004-6361/202039070} {\bibfield  {journal} {\bibinfo  {journal}
  {Astron. Astrophys.}\ }\textbf {\bibinfo {volume} {645}},\ \bibinfo {pages}
  {A104} (\bibinfo {year} {2021})},\ \Eprint {http://arxiv.org/abs/2007.15633}
  {arXiv:2007.15633 [astro-ph.CO]} \BibitemShut {NoStop}%
\end{thebibliography}%

\end{document}